\documentclass[fleqn,usenatbib]{mnras}

\usepackage{newtxtext,newtxmath}

\usepackage[T1]{fontenc}
\usepackage{ae,aecompl}

\usepackage{graphicx}	
\newcommand{\md}{{\mathrm{d}}}
\newcommand{\br}{{\bf r}}

\title[Cosmological nanolensing]{Cosmological nanolensing by dense gas clouds}

\author[Tuntsov and Walker]{Artem V. Tuntsov\thanks{E-mail: artem.tuntsov@manlyastrophysics.org},
Mark A. Walker\thanks{E-mail: mark.walker@manlyastrophysics.org}\\
$^{}$Manly Astrophysics, 15/41-42 East Esplanade, Manly, NSW 2095, Australia
}

\pubyear{2021}

\begin{document}
\label{firstpage}
\pagerange{\pageref{firstpage}--\pageref{lastpage}}
\maketitle

\begin{abstract}
We study the influence of a cosmological population of dense gas clouds on distant sources, with emphasis on quasar optical variability. In addition to gravitational lensing such clouds affect flux measurements via refraction in the neutral gas and via dust extinction, leading to a variety of possible light curves even in the low optical depth limit. We classify and illustrate the types of light curves that can arise. For sources as large as quasars we show that gravitational lensing and extinction are the dominant effects, with gas refraction playing only a minor r\^ole. We find that clouds with mass $\sim 10^{-4.5\pm 0.5}\,\mathrm{M}_\odot$ can reproduce the observed distribution of quasar variation amplitudes, but only if such clouds make up a large fraction of the closure density. In that case there may also be substantial extinction of distant optical sources, which can in principle be constrained by data on ``standard candles'' such as type Ia supernovae. Unfortunately that extinction is essentially grey, even when the material opacity is strongly wavelength dependent, making it difficult to distinguish from the influence of the background geometry. We propose a novel statistical test of the origin of quasar variability, based on the angular structure of the variation timescale for a large number of quasars distributed all over the sky. If quasar variability is primarily due to nanolensing that angular structure is expected to include a quadrupole term of amplitude $\sim5\%$, which ought to be measurable with future data from the {\it Gaia} mission.
\end{abstract}


\section{Introduction}
\label{section:intro}
There is now strong evidence for the existence of dark matter that far exceeds the amount of visible material on large scales \citep[e.g.,][]{1987ARA&A..25..425T, 2017NatAs...1E..59D}. Over the past three decades the idea that most of the dark matter is non-baryonic has become dominant, thanks mainly to the success of non-baryonic models in describing structure formation in the universe, as traced by galaxies and by the cosmic microwave background fluctuations \citep{2018ARA&A..56..435W,2020A&A...641A...1P}. Nevertheless the hypothesised ``dark matter particle'' remains elusive \citep{2010ARA&A..48..495F, 2018ARNPS..68..429B} and the picture thus lacks its most fundamental confirmation.

Even from that conventional perspective, however, baryonic matter is not fully accounted for either. Models of Big Bang Nucleosynthesis require twice as much baryonic matter as observed, in order to match the light-element abundances seen in the visible baryons \citep{1998ApJ...503..518F}. The ``missing baryons'' may perhaps be present in the form of a hot, ionised intergalactic medium \citep{1999ApJ...514....1C}, but the evidence to date has not yielded a precise measurement of the amount of this gas \citep[e.g.][]{2020Natur.581..391M}. Less  popular, but nevertheless interesting, is the idea that a lot of baryons may instead be locked up in dense, self-gravitating molecular clouds \citep{1994A&A...285...79P, 1995PhRvL..74...14D}. Such a population of undetected clouds would serve as a reservoir of material for ongoing star formation in galaxies \citep{1994A&A...285...79P}, and may be able to account for some of the observed regularities in those star-forming galaxies \citep{1999MNRAS.308..551W}.

Molecular clouds that are both cold and dense are very difficult to detect \citep[e.g.][]{1997A&A...327..453C, 1996ApJ...472...34G}. At low temperatures there is very little emission from molecular hydrogen itself, whereas the low vapour pressures of heavier, polar molecules such as CO limit their emissivity. Moreover, any microwave spectral lines that do contribute significantly to the emissivity of the gas are likely to be transitions into the ground state, which tend to be very optically thick -- because of the low temperature, high density and small Doppler broadening -- so radiation can escape a cloud only in the wings of those lines. Consequently we expect this material to fall into the category of ``CO-faint'' molecular gas \citep{2013ARA&A..51..207B}.

If such clouds were very dense indeed they could be compact enough to act as gravitational microlenses for sources in the Galactic Bulge or Magellanic Clouds \citep{1995ApJ...441...70H}; but in these settings it is also possible that they could be detected as a result of gas lensing --- i.e. refraction in the molecular gas itself \citep{draine1998}. If the gas is assumed to be completely transparent the Galactic microlensing data are able to exclude a substantial population in certain regions of the cloud mass-radius plane, but some mass-radius combinations remain largely unconstrained \citep{2001ApJ...547..207R}, depending on the assumed equation of state.

At the other extreme one can consider gas clouds that are totally opaque, but which have no lensing effect whatsoever.  \citet{2003ApJ...589..281D}, building on the predictions of \citet{2002MNRAS.332L..29K}, used data from the MACHO experiment \citep[][]{2000ApJ...542..281A} to search for extinction events due to Galactic clouds. They found no clear examples of extinction events and were thus able to exclude a dynamically significant, opaque population of clouds at the upper end of the planetary mass range. Opaque clouds with low masses ($\la 10^{-4}\,\mathrm{M}_\odot$) yield only short-lived extinction events and are not strongly constrained by the \citet{2003ApJ...589..281D} analysis.

More generally we expect the hypothesised cloud population to cause both refraction and extinction, but at present there are no reliable predictions of the form of the resulting light-curves. \citet{walkerwardle2019} presented detailed models of the internal structure of cold, dense clouds in which molecular hydrogen is condensing into solid or liquid form. Although detailed, those models are not yet accurate enough to allow us to construct reliable lightcurve templates. And without a suitable template the Galactic microlensing data of experiments such as OGLE \citep[][]{2000AcA....50....1U} cannot constrain the hypothesised population --- we do not know what we are looking for. Furthermore the fields targeted by the microlensing experiments tend to be chosen for their low mean extinction, and are therefore biased against finding dusty objects. This is in stark contrast to the situation with a population of primordial black holes, which are accurately modelled as point-mass gravitational lenses and whose abundance in the planetary mass range can be strongly constrained by the Galactic microlensing experiments \citep[e.g.][]{2020ARNPS..70..355C}. We note, however, that the constraints do relax somewhat if the mass spectrum is broad \cite{2021RPPh...84k6902C}.

If there is a wider cosmological distribution of dark gas clouds they would lead to variations in the received flux of distant, compact sources such as quasars. In this arena \cite{1993Natur.366..242H,1996MNRAS.278..787H} has long maintained that the observed variability of field quasars is best explained by a cosmologically significant population of gravitational microlenses, initially suggested to be\footnote{In later papers, e.g. \citet{2011MNRAS.415.2744H}, the same author suggested a lensing interpretation relying on stellar mass black holes instead. Stellar mass lenses were demonstrated to be a poor match to the data by the analysis of \cite{1993A&A...279....1S}, although those constraints do depend on the assumed model parameters, particularly the source size \cite{2003A&A...399...23Z}. We discuss quasar size estimates in an Appendix to this paper. Stellar mass lenses would also strongly magnify some type Ia supernovae at redshifts $z\sim1$ \citep[][]{1999ApJ...519L...1M}, and the lack of any such examples in existing data constrains the cosmological density of stellar mass lenses to be $\Omega < 0.1$ \citep[][]{2018PhRvL.121n1101Z}.} primordial black holes of mass $\sim 10^{-3}\,\mathrm{M}_\odot$. And in a similar vein \cite{1996ApJ...464..125S} inferred the presence of large numbers of $\sim 10^{-5}\,\mathrm{M}_\odot$ objects, which he dubbed ``rogue planets'', in the halo of the galaxy responsible for gravitationally lensing the multiply-imaged quasar Q$0957{+}561$.

By contrast, most of the recent work on multiply imaged quasars \citep[e.g.][]{2015ApJ...799..149J} has interpreted differences amongst the light curves of the various images as due to microlensing by \emph{stars} in the lens galaxy, with the dark matter assumed to be smoothly distributed on the relevant angular scales --- consistent with the dark matter being elementary particles rather than macroscopic bodies. Certainly the lens galaxies do contain stars, and those stars will cause microlensing, but it has been pointed out by \cite{2020A&A...633A.107H, 2020A&A...643A..10H} that the surface density of stars (at the macro-image locations) may in some cases be too low to account for the observed variations. Furthermore, if most of the microlenses are of stellar mass then the implied quasar sizes (radii) are larger than expected on the basis of accretion disk models \citep[e.g.][]{1973A&A....24..337S}. Source size is a key input parameter for modelling microlensing, and in Appendix A we give an overview of the available constraints on quasar sizes.

A clear theoretical perspective on the possibility of widespread microlenses was established long ago by \cite{1993A&A...279....1S}, who considered gravitational microlensing of quasars by point-mass objects randomly distributed between the Earth and the quasar. Schneider's results demonstrated that the observed variability of field quasars was not well explained by a substantial cosmological density of massive lenses ($M\ga 10^{-3}\,\mathrm{M}_\odot$), but left open the possibility of a microlensing interpretation utilising a large population of low-mass lenses ($M\la 10^{-5}\,\mathrm{M}_\odot$). These scenarios were revisited by \cite{2003A&A...399...23Z}, who showed that the constraints do depend on the assumed cosmology and source properties --- source size in particular (again, please see Appendix A of this paper for an overview of quasar size constraints). \cite{2003A&A...408...17Z} concluded that gravitational lensing by planetary mass objects could indeed reproduce a variety of quasar light curve statistics, but that it could not be the whole story and there had to be another, presumably intrinsic, variation mechanism.

In the present paper we study how quasar light curves are affected by a cosmological population of dense, molecular gas clouds of sub-stellar mass. Whereas stellar mass lenses at cosmological distances yield Einstein ring radii in the micro-arcsecond range, the much lower mass lenses that we consider yield structure on the nano-arcsecond scale; thus instead of ``microlenses'' we are dealing with ``nanolenses''. Section~\ref{section:phenomenology} identifies the important scales that determine the qualitative character of the combined effect of gravitational lensing, gas refraction and extinction, and demonstrates how to classify the resulting light curves. For quasar nanolensing we find that it is usually either gravitational lensing or extinction that  dominates what is seen, with gas lensing playing a lesser r\^ole. In Section~\ref{section:application} we compare our model statistics for source variability amplitude to data on quasar variability, finding that clouds with mass $\sim 10^{-4.5\pm 0.5}\,\mathrm{M}_\odot$ can provide a good fit to the observed distribution, providing they comprise a suitably large mean cosmological density, but our analysis is unable to constrain the cloud parameter space much further. In Section~\ref{section:grey} we turn our attention to the systematic influence of extinction on distant sources, demonstrating that the effect is essentially grey, and can be significant for sources at redshifts $z\ga 1$. Grey extinction will be difficult to identify in practice, but in Section \ref{section:quadrupole} we present a new and purely kinematic effect -- quadrupolar modulation of variation timescale across the sky -- which will permit a test of the extrinsic origin of quasar variability.  Busy readers can skip straight to Section~\ref{section:conclusions} where our results are summarised.

\section{Phenomenology of gas cloud lensing}
\label{section:phenomenology}

\subsection{Gas clouds: gravitating, refractive and opaque}
\label{subsection:threeaspects}
There are three distinct physical effects that arise when radiation propagates through a population of molecular gas clouds: in addition to acting as a gravitational lens, a compact molecular cloud affects radiation via refraction in the neutral gas, and through extinction.

\subsubsection{Refraction}
Just like its gravitational counterpart, in the case of small deflections, as relevant for astrophysical applications, gas lensing can be described in terms of the phase delay imparted to each ray. That phase can be computed by integrating along the undeflected path:
\begin{eqnarray}
\Phi_g(\br)=\int\md z\, k\left[n(\br, z)-1\right].
\end{eqnarray}
Here $k$ is the wavevector and $n(\br, z)$ is the refractive index at the position $z$ along, and $\br$ across, a fiducial optical axis. The refractive index is just a linear function of the gas density, $\rho$, with $n-1=a_\lambda\rho$ \citep{draine1998}; we refer to the constant of proportionality, $a_\lambda$, as the specific refractivity.

It is possible for the gas to have a non-uniform composition. That is the case, for example, in the models of \citet{walkerwardle2019}, where precipitation of molecular hydrogen condensates makes the central regions rich in H$_2$, while the surface layers are almost pure helium. Such configurations have a specific refractivity that is not uniform. For the sake of simplicity in this paper we will consider only clouds of uniform composition, so that the phase delay is proportional to the column density $\Sigma(\br)$,
\begin{eqnarray}
\Phi_g(\br)=k a_\lambda \Sigma(\br).
\end{eqnarray}
For a mix of molecular hydrogen and helium, with a `primordial' He fraction of 24 per cent by mass, \cite{draine1998} gives a value 
\begin{equation}
a_\lambda\approx1.23\,\mathrm{cm}^3\mathrm{g}^{-1},
\end{equation}
with less than a 3 percent variation across the wavelength range from $440\,\mathrm{nm}$ to $670\,\mathrm{nm}$.

In application across the full electromagnetic spectrum a much larger variation in $a_\lambda$ would of course be encountered. At very long wavelengths the refractivity asymptotes to a value ($1.18\,\mathrm{cm}^3\mathrm{g}^{-1}$) only a few percent smaller than in the optical. Moving to shorter wavelengths the refractivity rapidly increases as the Lyman and Werner bands of ${\rm H}_2$ are approached, where it fluctuates rapidly atop a systematic decline; it is negligible in the X-ray band and above.

Note that gas lensing is local -- i.e., light rays are only affected by the matter they pass through. In contrast, the delay from gravitational lensing is due to a long-range force and can be related to the column density via a Poisson equation \cite[e.g.][]{sef1992}
\begin{eqnarray}
\nabla^2_\br\Phi_l(\br)=\frac{8\pi G k}{c^2}\Sigma(\br).
\end{eqnarray}
Appropriate boundary conditions ensure that a homogeneous mass distribution, being absorbed into the definition of angular diameter distances, does not contribute to light ray deflection. 

The total phase delay, in the limit of weak gravity and small deflections that we consider, is the sum of the gaseous and gravitational contributions,
\begin{eqnarray}
\Phi(\br)=\Phi_g(\br)+\Phi_l(\br).
\end{eqnarray}
We restrict ourselves to geometric (ray) optics, as appropriate in the optical, and assume that the lensing matter is concentrated in a region that is small along the line of sight compared to the angular diameter distances to the source $D_s$, to the lens $D_l$ and from the lens to the source~$D_{ls}$. In this case both gas and gravitational lensing act to bend the light rays by an angle
\begin{eqnarray}\label{physicalangle}
\hat\balpha(\br)=k^{-1}\nabla_\br\Phi(\br)
\end{eqnarray}
and the observed position $\pmb{\theta}$ of the image is related to the true position $\pmb\beta$ of the source via the lens equation
\begin{eqnarray}
\bbeta=\btheta-\balpha(\btheta).
\end{eqnarray}
The scaled deflection angle as a function of image position is related to the physical deflection on the lens plane~(\ref{physicalangle}) via
\begin{eqnarray}
\balpha(\btheta)=\frac{D_{ls}}{D_s}\hat\balpha(D_l\btheta).
\end{eqnarray}
Apart from extinction, which we turn to in a moment, and redshift, which is constant on the timescales of interest to us, intensity is conserved along rays, hence variations in the observed flux are purely due to changes in the solid angle of the image compared to that of the source. Locally -- i.e., for a small source -- the magnification is thus given by the Jacobian of the lens mapping:
\begin{eqnarray}
\mu=\det\partial\theta_i/\partial\beta_j=\frac1{(1-\varkappa)^2-\gamma^2}
\end{eqnarray}
where convergence, $\varkappa$, and shear, $\gamma$, have been introduced:
\begin{eqnarray}
\varkappa=\frac{D_{ls}D_l}{2kD_s}\left(\partial^2_x\Phi+\partial^2_y\Phi\right),\\
\gamma=\frac{D_{ls}D_l}{2kD_s}\sqrt{\left(\partial^2_x\Phi-\partial^2_y\Phi\right)^2+4\left(\partial^2_{xy}\Phi\right)^2}.
\end{eqnarray}
The convergence and shear define eigenvalues $1-\varkappa\mp\gamma$ of the transformation matrix $\partial\beta_j/\partial\theta_i$ that are independent of the coordinate system. The convergence due to the combination of gas and gravitational lensing is simply the sum of the two individual convergences and, remarkably, it is local. The local character of the gas refraction is obvious; in the case of the gravitational contribution it follows from the Poisson equation and we have
\begin{eqnarray}
\varkappa=\left( 1 +\ell_\lambda^2 \nabla^2 \right) \frac\Sigma{\Sigma_{cr}}.
\end{eqnarray}
In this result for the beam convergence, which is dimensionless, there are two characteristic scales. The first is the familiar one for gravitational lensing, namely the critical surface density
\begin{eqnarray}\label{criticaldensity}
\Sigma_{cr}=\frac{c^2 D_s}{4\pi G D_l D_{ls}}\approx3.475\times10^2\,\mathrm{g}\,\mathrm{cm}^{-2}\left(\frac{D_lD_{ls}/D_s}{1\,\mathrm{Mpc}}\right)^{-1},
\end{eqnarray}
which a gravitational lens must exceed in order to produce multiple images. In the absence of any gas lensing ($\ell_\lambda=0$) the beam convergence is just that due to gravitational lensing and its value is simply the value of the column-density, $\Sigma$, relative to the critical value. 
The second characteristic scale in equation (12) is a length scale: the gas lensing curvature radius, $\ell_\lambda$, given by
\begin{eqnarray}\label{gasscale}
\ell^2_\lambda=\frac{a_\lambda c^2}{4\pi G}\approx\left(3.6\times10^{13}\,\mathrm{cm}\right)^2.
\end{eqnarray}
Here the numerical estimate corresponds to the particular value of specific refractivity given in equation (3). This length scale  plays a key r\^ole because it determines the strength of the gas lens relative to that of the gravitational lens. If the cloud has a column-density curvature radius that is small (large) compared to $\ell_\lambda$ then the beam convergence due to gas lensing will be large (small) compared to that due to gravitational lensing.

Magnification of a source of uniform brightness is, as usual, just the ratio of the total solid angle of all images to that of the unlensed source. Given that gas lensing is local, it cannot lead to the formation of images outside the radius of the cloud, and it can only lead to significant magnification for sources whose angular size is much smaller than that of the cloud. In instances where gravitational refraction leads to image formation beyond the limb of the cloud the properties of those images are unaffected by gas lensing.

Provided that the cloud surface-density doesn't greatly exceed the mean column, $\langle \Sigma\rangle\equiv M/\pi R^2$ for cloud mass $M$ and radius $R$, equation (12) allows us to estimate the importance of gas lensing and gravitational lensing, as follows. The Einstein ring radius, $R_E$, which is the radius at which the images form for a pure, strong gravitational lens with a source on-axis, is given by $\pi R_E^2 = M/\Sigma_{cr}$. Thus with $\Sigma\sim \langle \Sigma\rangle$ the gravitational contribution to equation (12) -- i.e. the first term in parentheses on the right-hand side -- is  $\sim (R_E/R)^2$. The contribution of gas lensing, on the other hand -- i.e. the second term in parentheses on the right-hand side -- is $\sim (\ell_\lambda R_E)^2/R^4$. Thus we see that, just as $R_E$ is a critical cloud radius for gravitational lensing, the critical radius for gas lensing is the geometric mean of $R_E$ and $\ell_\lambda$.

\begin{figure}
\includegraphics[width=0.95\columnwidth]{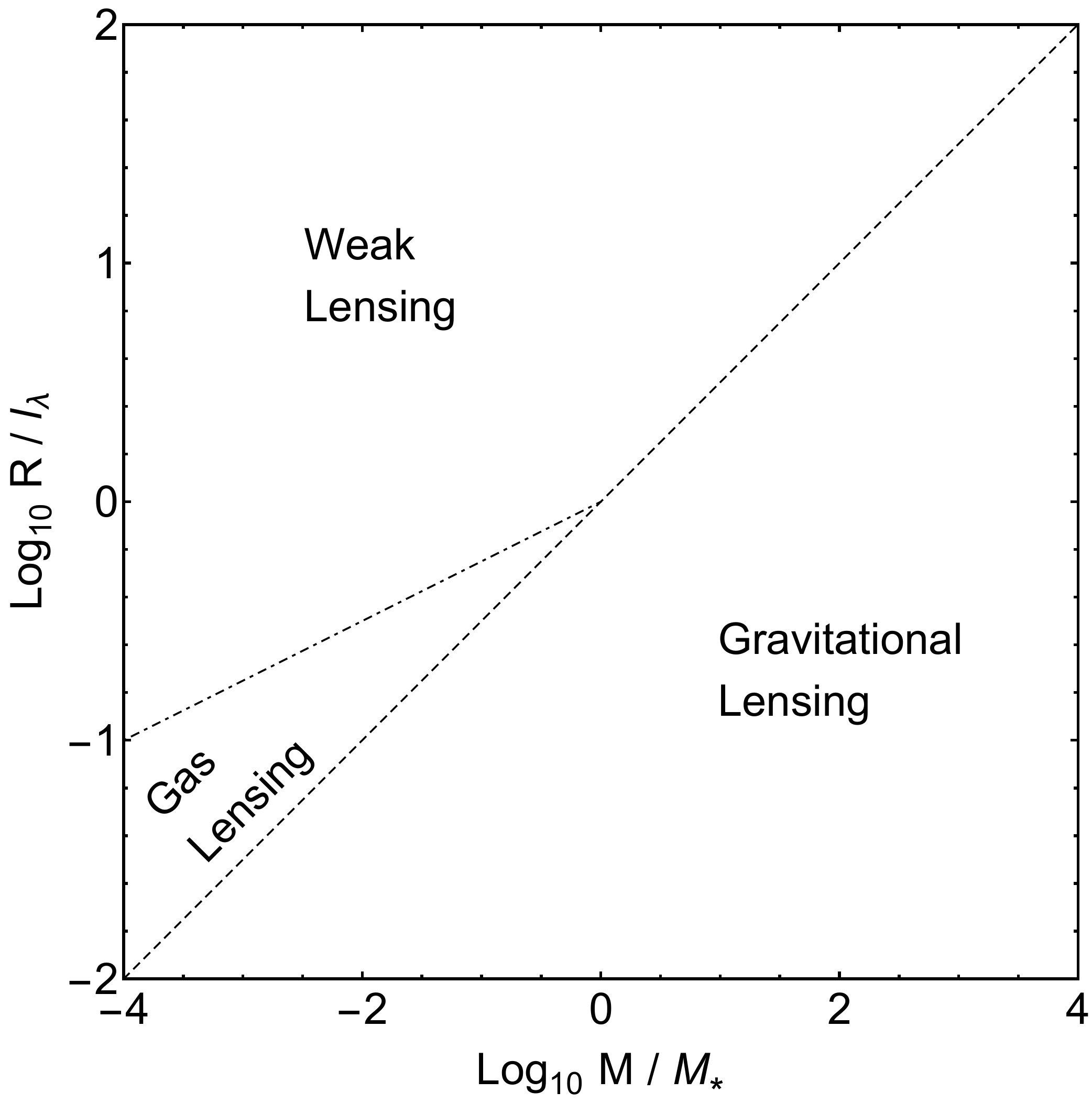}
\caption{Classification of lens types according to their position in the mass-radius plane. In this plot radii are in units of the gas-lensing curvature scale, $\ell_\lambda$, and masses are in units of the fiducial value $M_*=\pi \ell_\lambda^2\,\Sigma_{cr}$, for which the Einstein radius $R_E=\ell_\lambda$. For any given optical configuration the transition to a strong gravitational lens corresponds to a particular surface density, $\Sigma_{cr}$, for which the critical radius is $R=R_E\propto \sqrt{M}$. The critical radius for gas lensing is $\sqrt{R_E \ell_\lambda} \propto M^{1/4}$, and is only important for $M<M_*$.}
    \label{figure:lenstypes}
\end{figure}

The existence of a critical column-density for gravitational lensing, $\Sigma_{cr}$  (equation 13), and the curvature scale for gas lensing, $\ell_\lambda$ (equation 14), together imply a fiducial cloud mass, $M_*\equiv \pi \ell_\lambda^2 \Sigma_{cr}$. And the point $(M_*,\ell_\lambda)$ in the mass-radius plane lies at the intersection of the boundaries of three different regimes of lens behaviour, as shown in figure \ref{figure:lenstypes}. In that figure the upper boundary of the region of strong gravitational lensing is simply a line of constant column-density -- the critical surface density, $\Sigma_{cr}$, given in equation (13) -- i.e. $(M/M_*)=(R/\ell_\lambda)^2$. Equivalently we can describe the upper boundary of strong gravitational lensing by the condition $R=R_E$ (the Einstein ring radius). As noted above, the corresponding critical radius for strong gas lensing is $R=\sqrt{R_E\,\ell_\lambda}$, and that boundary can be written in the alternative form $(M/M_*)=(R/\ell_\lambda)^4$. We note that the boundary of the region of strong gas lensing is only shown for $R<\ell_\lambda$ in figure 1, for the following reason. Clouds that are larger than $\ell_\lambda$ exhibit more gravitational refraction than gas refraction, so if gas lensing is strong and $R>\ell_\lambda$ then gravitational lensing is guaranteed to be even stronger and the main images will form outside the cloud (where gas refraction is irrelevant).

Because the critical radius for gas lensing can be much larger than the Einstein ring radius, if $\ell_\lambda \gg R_E$, it follows that for some types of clouds the optical depth to gas lensing can be much larger than the optical depth to gravitational lensing -- by a factor $\ell_\lambda/R_E \gg 1$ -- with correspondingly high event rates predicted for microlensing of stars in the Galaxy \citep[][]{draine1998,2001ApJ...547..207R}.

The fiducial mass just discussed depends on line-of-sight distances, but is independent of the properties of the source. However, in order for a lens to be able to substantially magnify a source we require that the lens must be larger than the (projected size of the) source. Therefore, for typical lenses with $D_l\sim D_{ls}\sim D_s/2$, the gas lensing curvature scale itself also demarcates two distinct regimes of behaviour: for large sources with $R_s\gg 2 \ell_\lambda$ it is \emph{only} possible to obtain substantial magnification with gravitational lensing, and then only with clouds that are more massive than the fiducial value $M_*$. The two types of sources that we are concerned with here -- quasars and supernovae -- both have radii $R_s\gg 2 \ell_\lambda$, and it is for that reason that gas lensing is of little consequence in this paper.

We have already noted the wavelength dependence inherent in neutral gas refraction, but in this paper we will not explore that dimension. Instead we will concentrate on describing the behaviour in the optical (specifically the $\mathrm{V}$-band), supplemented by some brief commentary on how the situation changes across the broader spectrum. Even with that restriction we demonstrate a wide variety of behaviour in the model. We note in passing that measurements made in $\mathrm{V}$-band would correspond to shorter wavelengths -- hence a larger refractivity -- at the location of a cosmologically distant cloud where the lensing actually occurs. A similar point applies to the extinction introduced by a cloud.

\subsubsection{Extinction}
We expect there to be two distinct contributions to the extinction of light: absorption and scattering by the individual atoms and molecules that make up the gas, and the extinction due to any dust particles that are present. Helium atoms and hydrogen molecules both respond strongly in the far-UV, and only comparatively weakly elsewhere. Below that band the larger contribution by far comes from H$_2$. The Rayleigh scattering cross-section for H$_2$ is a strong function of wavelength ($\lambda^{-4}$), and in the $\mathrm{V}$-band it contributes an opacity of \citep[][]{1962ApJ...136..690D} $2.1\times10^{-4}\,\mathrm{cm}^{2}\,\mathrm{g}^{-1}$ (after allowance is made for the mass contribution of the helium). In the optical the hydrogen molecule has some weak (quadrupole) rovibrational absorption lines, but in any broad-band filter the absorption would be dominated by the low-frequency wings of the far-UV resonance lines. This absorption can be modelled as a sum of Lorentz oscillators, as described by \citet{kettwich2015}. With that model the expected absorption opacity is $1.1\times10^{-3}\,\mathrm{cm}^{2}\,\mathrm{g}^{-1}$, in $\mathrm{V}$-band, varying approximately as $\lambda^{-2}$. The resulting optical depth to extinction, $\tau_\lambda$, can be written as
\begin{eqnarray}
\tau_\lambda(\br)=\kappa_\lambda\Sigma(\br),
\end{eqnarray}
where $\Sigma(\br)$ is the total gas column, and $\kappa_\lambda$ is the total opacity (absorption plus scattering) of the cloud material, which we will often quote the inverse of, $\kappa_\lambda^{-1}$ for ease of comparison with the other column density scales of the problem; as an inverse of the opacity, this will be referred to as a material transparency. The molecular contribution to opacity is unavoidable and implies an upper limit on $\kappa_\lambda^{-1}$ of $768\,\mathrm{g}\,\mathrm{cm}^{-2}$ in $\mathrm{V}$-band.

We have already noted that in the far-UV the atomic and molecular transparencies are very low. At much shorter wavelengths still the absorption becomes negligible and one is left with only the electron scattering (Thomson), which amounts to $\kappa_{\lambda,\mathrm{T}}^{-1}\approx 2.9\;\mathrm{g}\,\mathrm{cm}^{-2}$ for gas which is $76$\%\ hydrogen and $24$\%\ helium, by mass. At the other extreme of wavelengths that are much longer than the optical the transparency increases rapidly; opacity is negligible in the radio.

The total opacity in the optical could be much higher than the molecular value, as a result of extinction by dust, but the contribution of the dust is uncertain even to order of magnitude. Consequently there are combinations of cloud mass and radius for which some physically plausible models would imply that essentially all the incident light is absorbed, and other models in which essentially all incident light is transmitted. To deal with the uncertainty introduced by the unknown dust opacity we proceed by bracketing the range of possibilities; we do this in two distinct, alternate ways --- depending on which seems most appropriate. When considering clouds whose mass and radius are specified only in terms of $M_*$ and $\ell_\lambda$, respectively, the column-density in ${\rm g\,cm^{-2}}$, say, is not specified and so it is inappropriate to work with specific values of the opacity. In that case we specify the central optical depth to extinction and we consider the limiting cases of totally transparent and totally opaque clouds, as shown in figure 3. On the other hand, when considering clouds in a specific context we bracket the range of possible opacities in the manner described below.

For simplicity, we assume that equation (15) also describes the extinction due to dust particles, with a single, suitably chosen value of $\kappa_\lambda$ that is constant throughout each cloud and across the whole cloud population. In the local ISM the coefficient of proportionality between total hydrogen column and $\mathrm{V}$-band extinction is measured to be $N_\mathrm{H}/A_\mathrm{V}\approx1.9\times10^{21}\,\mathrm{cm}^{-2}\mathrm{mag}^{-1}$
\citep{bohlin1978}, which corresponds to 
\begin{equation}
\kappa_{\lambda=V}^{-1}\approx3.6\times10^{-3}\,\mathrm{g}\,\mathrm{cm}^{-2}. 
\end{equation}

The aforementioned numbers apply to the local, diffuse ISM and might not be representative of the material in cold, dense, self-gravitating gas clouds. In particular we note that self-gravitating clouds could in principle be much more transparent than the local diffuse ISM because any dust particles made of refractory elements could  have sedimented into a dense core \citep{draine1998}. Even in that case, however, hydrogen condensates may form if the gas is both cold and dense \citep[][]{1994A&A...285...94P}. Condensed hydrogen has very little absorption in the optical band \citep{kettwich2015}, but small particles of hydrogen ``dust'' would scatter optical light and thus cause extinction. To gauge the level of extinction that might arise from hydrogen condensates we turn to the hydrogen ``snow cloud'' models of \citet{walkerwardle2019}. In those models the ``snowflake'' content of the gas is highly non-uniform, being zero in the cloud core -- which is too warm to permit hydrogen condensation -- and near zero in the outer layers, which are almost pure helium, but substantial at intermediate radii. Nevertheless we can see from panel (c) of figure 3 in \citet{walkerwardle2019} that, for the particular cloud shown there, the column density in hydrogen snowflakes ought to be roughly a fraction $\sim 10^{-3}$ of the total column, for impact parameters below $2.4\;{\rm AU}$. The column-density of a single, micron-sized particle of solid hydrogen is $\sim 10^{-5}\,\mathrm{g}\,\mathrm{cm}^{-2}$, so if we assume all the condensate to be in that form then we arrive at a fiducial $\kappa_{\lambda}^{-1}\sim 10^{-2}\,\mathrm{g}\,\mathrm{cm}^{-2}$ --- an estimate that is not far from the value appropriate to the local ISM.

Unfortunately \citet{walkerwardle2019} did not attempt to model the size of the condensed particles so we do not have a value for their characteristic column-density. An upper limit can be estimated by noting that particles with very large column-densities are removed from the atmosphere on a dynamical timescale, because gravity is then the dominant force, whereas small particles can be kept in suspension by the ram pressure associated with convective updraughts. If we assume that the latter have a speed that is of order one tenth of the sound-speed in the gas, balancing ram pressure with gravity at the inner-edge of the condensing envelope of the cloud in figure 3 of \citet{walkerwardle2019}, leads to a characteristic column-density $\sim 1\,\mathrm{g}\,\mathrm{cm}^{-2}$ for the condensed particle --- corresponding to a sphere of radius $10\;\mathrm{cm}$. It is therefore possible that the transparency due to snowflakes could be as high as $\kappa_\lambda^{-1}\sim 10^{3}\,\mathrm{g}\,\mathrm{cm}^{-2}$, which is comparable to the molecular contribution. 

Given the large uncertainties in the transparency that is due to dust, the best we can do is to bracket the opaque column within a broad range: 
\begin{equation}\label{opacityrange}
768\,\mathrm{g}\,\mathrm{cm}^{-2}\quad \ga \quad \kappa_{\lambda}^{-1} \quad \ga \quad 3.6\times10^{-3}\,\mathrm{g}\,\mathrm{cm}^{-2}
\end{equation}
(in $\mathrm{V}$-band).

As with gas lensing, extinction is local so only images formed inside the cloud are affected: if gravitational refraction leads to image formation beyond the limb of the cloud then both gas lensing and extinction are irrelevant for that image. There is thus a regime where the effect of a compact molecular gas cloud resembles that of a point-mass gravitational lens, effecting a major simplification in the light curve structure (see \S\ref{subsection:lcclassification}). Conversely, if the clouds are sufficiently nearby that their column-density is below the critical surface density (13) then the dominant image can form within the projected radius of the cloud and it is possible for transient extinction events to arise. Notwithstanding this broad-brush picture, the detailed structure of the light curves is affected to some degree by both extinction and gas lensing, even for distant/compact lenses --- through the extinction of sub-dominant images, for example.

\subsection{Gaussian model cloud profile}
\label{subsection:models}
We utilise an axisymmetric gaussian model column-density profile for the individual clouds, to illustrate lensing plus extinction effects. A gaussian profile is convenient because it is amenable to analytic treatment and consequently has been often used to model discrete clouds \citep[e.g.][]{1998ApJ...496..253C,1999MNRAS.306..504W}. The model projected density is
\label{subsubsection:gaussianlens}
\begin{eqnarray}\label{gaussianlens}
\Sigma({\bf r})=\Sigma_0\exp\left[-\left(\frac{r}{r_0}\right)^2\right],
\end{eqnarray}
which has two free parameters, the central column density $\Sigma_0$ and lens scale $r_0$.

This leads to the lens equation between the radial position of the source $y$ and its image $x$ relative to the centre of the lens:
\begin{eqnarray}\label{glensequation}
y=x-s\left(\frac{1-\mathrm{e}^{-x^2}}x+2gx\mathrm{e}^{-x^2}\right),
\end{eqnarray}
with positive (negative) $x$ being interpreted as images lying on the same (opposite) side of the optical axis as the source. Both $x$ and $y$ are quantities projected onto the lens plane, so that the effective size of the source is scaled by the angular diameter distance ratio $D_{l}/D_{s}$, and expressed in units of the lens scale $r_0$ (or its angular equivalent $\theta_0=r_0/D_{l}$). The two free parameters have been recast as the central column density in critical units~(\ref{criticaldensity}), and the (square of the) gas lensing curvature scale~(\ref{gasscale}) in units of the lens radius:
\begin{eqnarray}
s\equiv\frac{\Sigma_0}{\Sigma_{cr}}, \hspace{.5cm} g\equiv\left(\frac{\ell_\lambda}{r_0}\right)^2=\left(\frac{\ell_\lambda}{D_{l}\theta_0}\right)^2.
\end{eqnarray}

For $s(1+2g)>1$ there are two critical curves and their corresponding caustics where the magnification factor diverges. One is the tangential critical curve (Einstein ring) that corresponds to a non-trivial solution of the equation (19) at $y=0$, hence the corresponding caustic lies on the optical axis. The other is the radial critical curve that is the position $x$ where $\mathrm{d}y/\mathrm{d}x$ vanishes, and the corresponding position $y$ in the source plane is the radial caustic. Far from the lens, $y\gg\sqrt{s(1+2g)}$, the mapping is nearly the identity and there is only one solution, $x\approx y$. If the source moves towards the lens and crosses the radial caustic, two additional images appear either side of the radial critical curve. As the source is brought closer to the optical axis the two extra images move away from the radial critical curve, one towards the lens centre and the other towards the Einstein ring where it merges with the primary image as $y\to0$.  For $s(1+2g)<1$ the mapping is one-to-one throughout. 

Equation~(\ref{glensequation}) has no close-form solution and we use Newton's method to solve it. In the case of the primary image we start our iteration from the point $x=y$. For any additional images, $y$ is inside the radial caustic, we start the iteration from halfway between the radial critical curve and either the lens centre or the tangential critical curve. The positions of the caustic and the radial curves are not available as close-form expressions either and are likewise computed, for given $g$ and $s$, using Newton's method.

Depending on the size of the source relative to the radial caustic (if present), and its position, all or a part of the source may be multiply imaged with the extra images possibly merging at the radial critical curve. We model the source as a disc of uniform brightness, so that the total flux can be readily computed once the limb of the source is mapped into the image plane (thus avoiding a two-dimensional averaging of the magnification). To proceed we break the source into a number of small, quasi-rectangular patches by chords that are radial with respect to the centre of the lens, map their end points, and count the total area of the resulting images. Extinction depends on the position of the image and, unlike gravitational and gas lensing, modifies the brightness. The introduction of extinction necessitates integration along each chord image, rather than just mapping the boundary positions, but in the particular case of a gaussian column-density profile the result can be expressed analytically in terms of the exponential integral.

\begin{figure}
    \centering
    \includegraphics[width=0.95 \columnwidth]{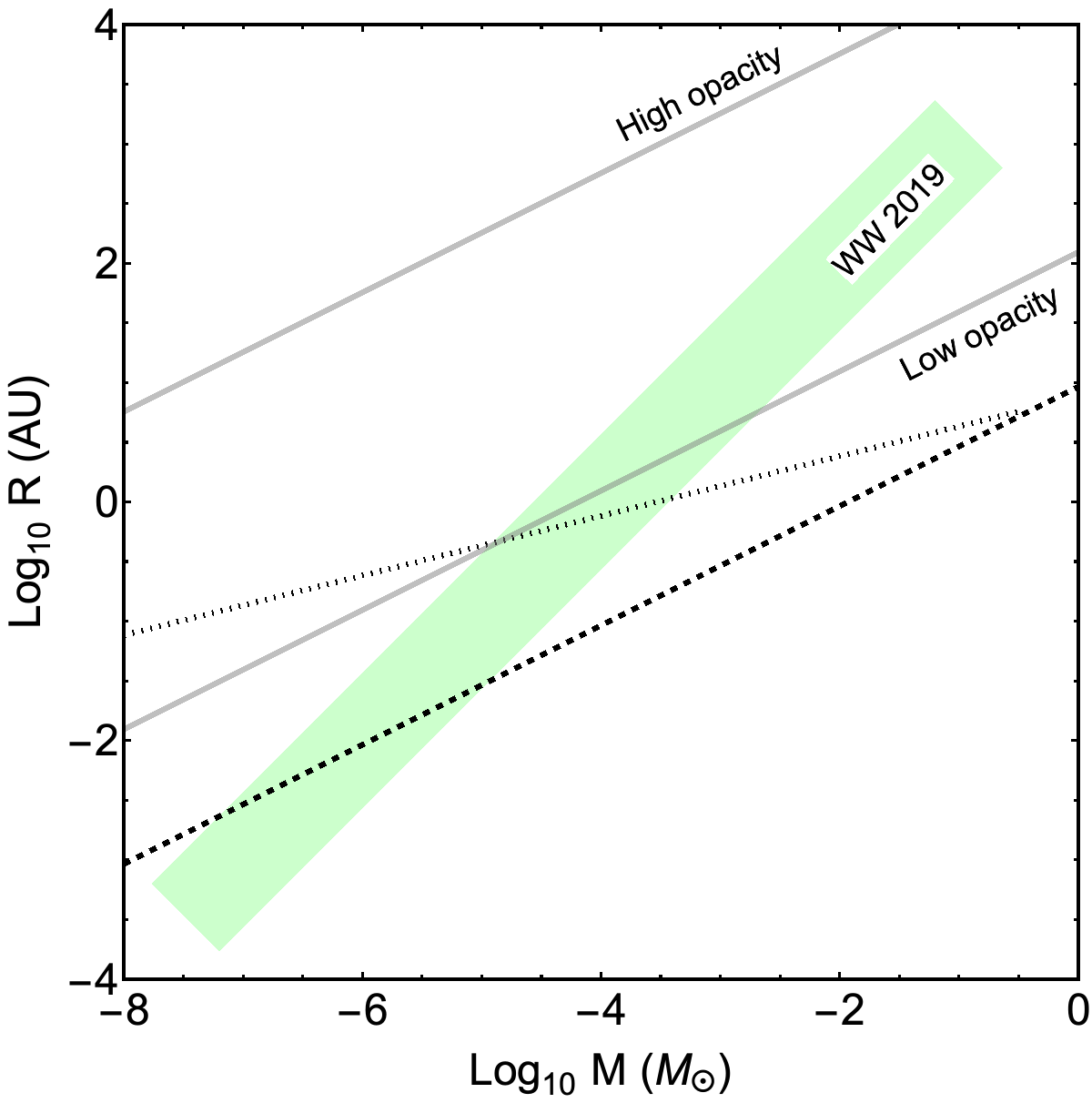}
     \includegraphics[width=0.95 \columnwidth]{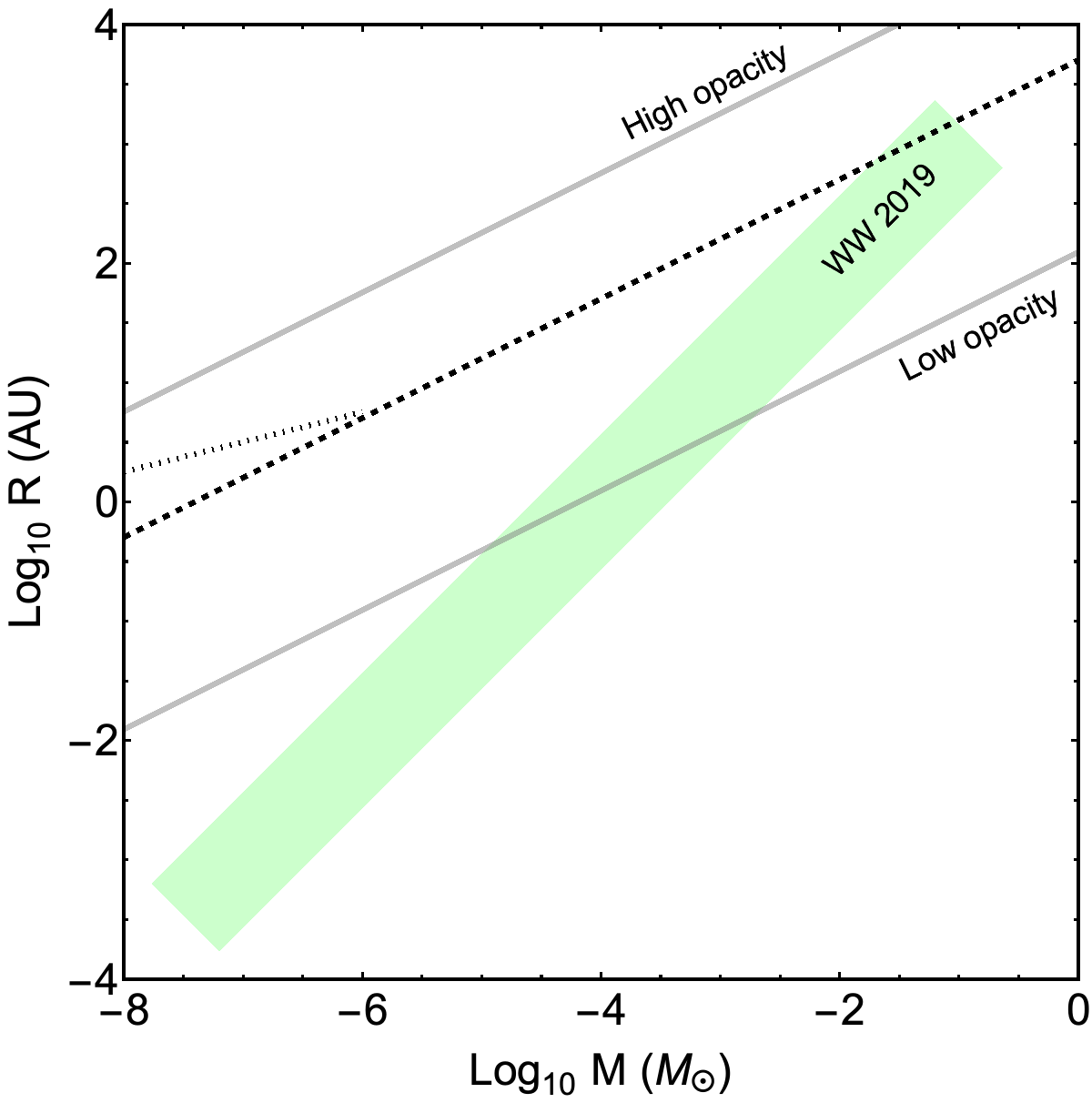}
    \caption{Partitioning of the mass-radius plane for molecular gas clouds according to their lensing and extinction characteristics. The upper (lower) panel is for Galactic (cosmological) clouds. The green shaded area roughly delineates the ``snow-cloud'' models shown in figure 4 of \citet{walkerwardle2019}. Dashed lines show the transition to a strong gravitational lens for $n=3/2$ polytropes; dotted lines show the corresponding transition to strong gas lensing. The lower (upper), solid, grey line corresponds to a column $\Sigma=\kappa_\lambda^{-1}$ for the maximum (minimum) material transparency considered in \S2.1.2. Clouds lying to the left (right) of the appropriate transparency line have a column-density below (above) the opaque column, $\kappa_\lambda^{-1}$, and thus are transparent (opaque).}
    \label{figure:massradius}
\end{figure}

\subsection{Light-curve classification for point-like sources}\label{subsection:lcclassification}
If we want to know the details of the light-curves that a given gas cloud might generate then we need to undertake calculations specific to that case, but once the properties of the clouds are specified the qualitative nature of the light-curves can be anticipated without detailed calculations  --- as we now describe. We first consider the lensing behaviour and then turn to the influence of opacity.

\subsubsection{Position in the mass-radius plane}
Typically one is considering hypothetical cloud models that have a well-defined mass and radius, and thus can be placed in a diagram such as figure 1. The type of lensing light-curves that could result from occultation of a source is then determined by which sector the cloud resides in --- although, of course, the classification is unambiguous only when the cloud is located far from the sector boundaries.

In the Weak Lensing domain there is no significant refraction, hence no noticeable change in flux due to lensing. In the Gravitational Lensing domain the influence of the cloud is similar to that of a point-like mass, and the light-curve will be close to a Paczy\'nski curve \citep[][]{1986ApJ...304....1P}. In the Gas Lensing domain there is no unique functional form for the light-curves, but the generic structure of the flux maps is a demagnification region where the source is behind the limb of the cloud and a magnification region where it is on-axis. Examples of gas lensing behaviour can be found in the next section for the specific case of clouds having a gaussian column-density profile, and in the literature \citep[][]{draine1998,2001ApJ...547..207R} for the case of polytropic clouds.

If nothing more is known about the cloud than its mass and radius then figure 1 itself should be used for classification, but we often have further information about the physics of the structures under consideration and in that case a clearer differentiation can be achieved by using a more precise construction. For example the gas cloud models shown in figure 4 of \citet{walkerwardle2019} are not far from $n=3/2$ polytropes, because they comprise a large core that is precisely polytropic and only a small, non-polytropic envelope. For any pure polytropic model the relationships between point-wise properties and their average values can be quantified, so we can determine values for both contributions to the beam convergence in equation (12) in terms of the cloud mass and radius. 

The gravitational lensing beam convergence is simply $\Sigma/\Sigma_{cr}$, and for $n=3/2$ polytropes the peak (central) value of the column-density is $\Sigma_{max}\simeq 4.15\,\langle \Sigma \rangle = 4.15\, M/(\pi R^2)$; thus in this case the transition to a strong gravitational lens occurs at a cloud radius that is approximately twice as large as for gravitational lensing by the cloud as a whole. Polytropic clouds with $n>3/2$ can show much larger differences in that respect, because their surface density profiles are increasingly centrally peaked as $n$ increases.

Similar considerations apply to the onset of strong gas lensing. For polytropic clouds with $n=1.5, 2.0, 2.5, \dots , 4.5$ that transition was quantified by \citet{draine1998} as a gas lens strength $S=S_c(n)$, i.e. a critical value of the strength parameter, depending on the polytropic index. That description can be easily related to our parameterisation, because Draine's lens strength is $S=(3/4)(M/M_*)(\ell_\lambda/R)^4$. The critical value for $n=3/2$ is $S_c(n=3/2)\simeq0.026$ \citep{draine1998}, and decreasing rapidly with increasing $n$. 

\begin{figure*}
\centering
{\includegraphics[width=0.3\textwidth]{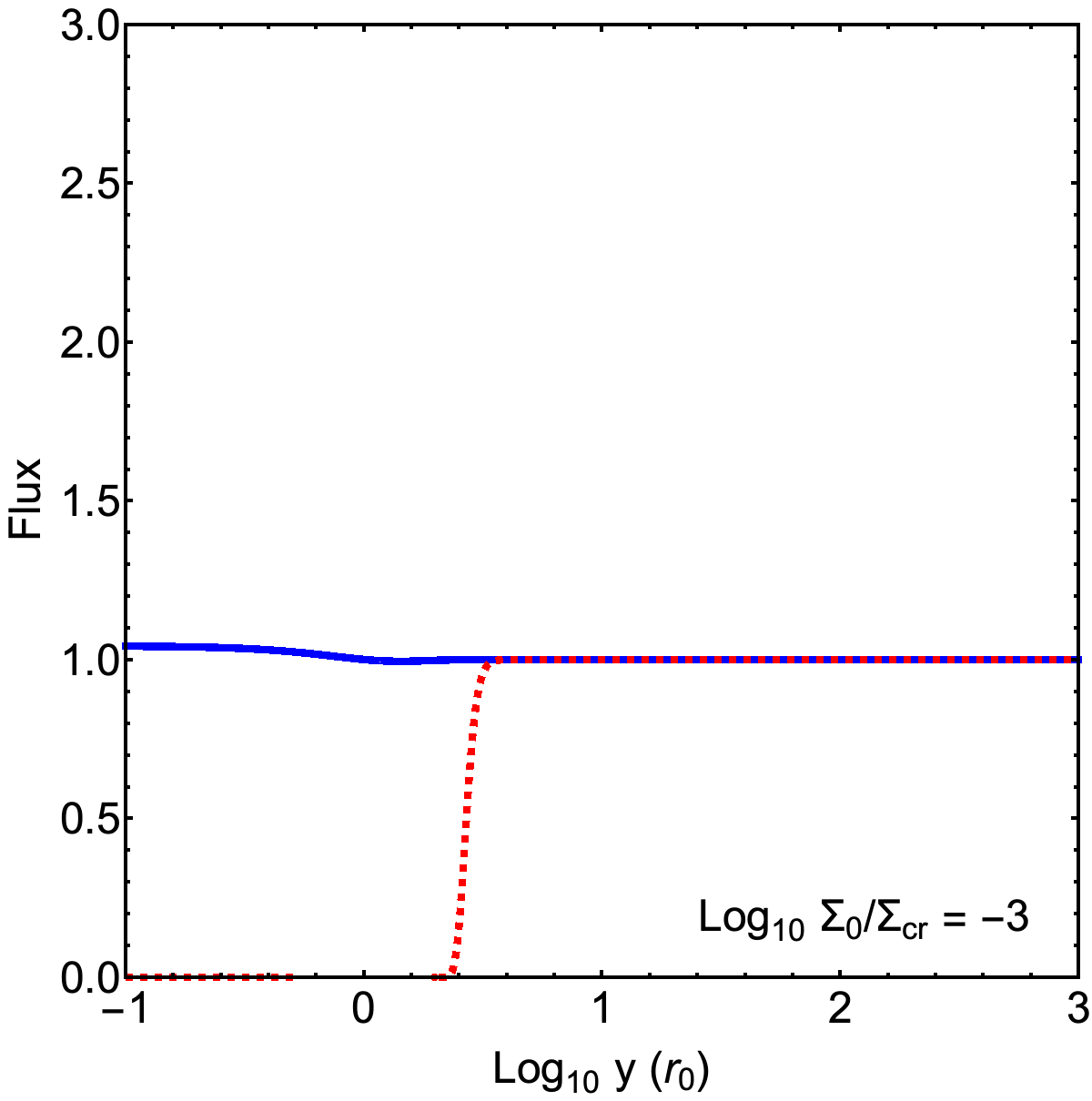}
\includegraphics[width=0.3\textwidth]{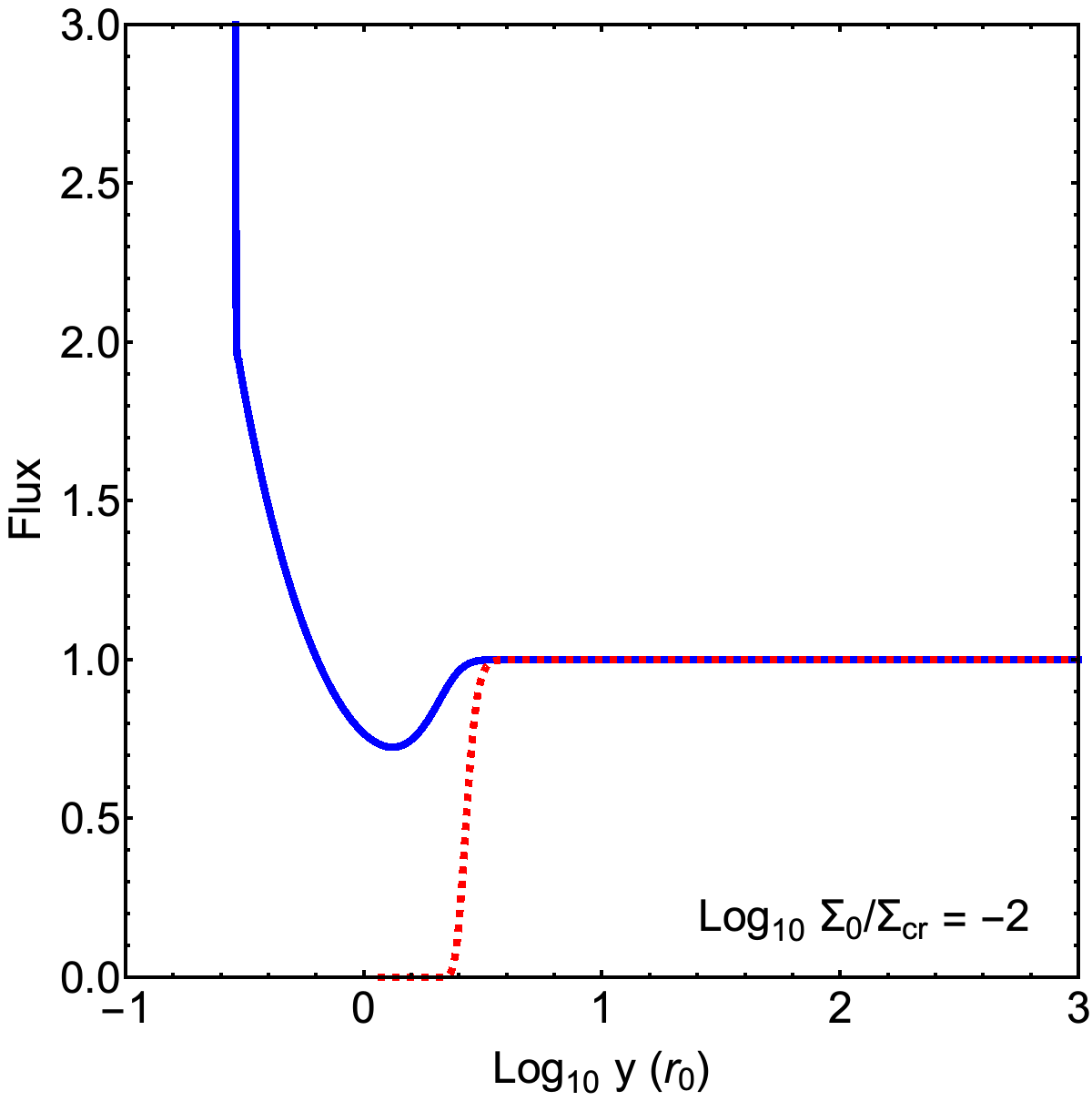}
\includegraphics[width=0.3\textwidth]{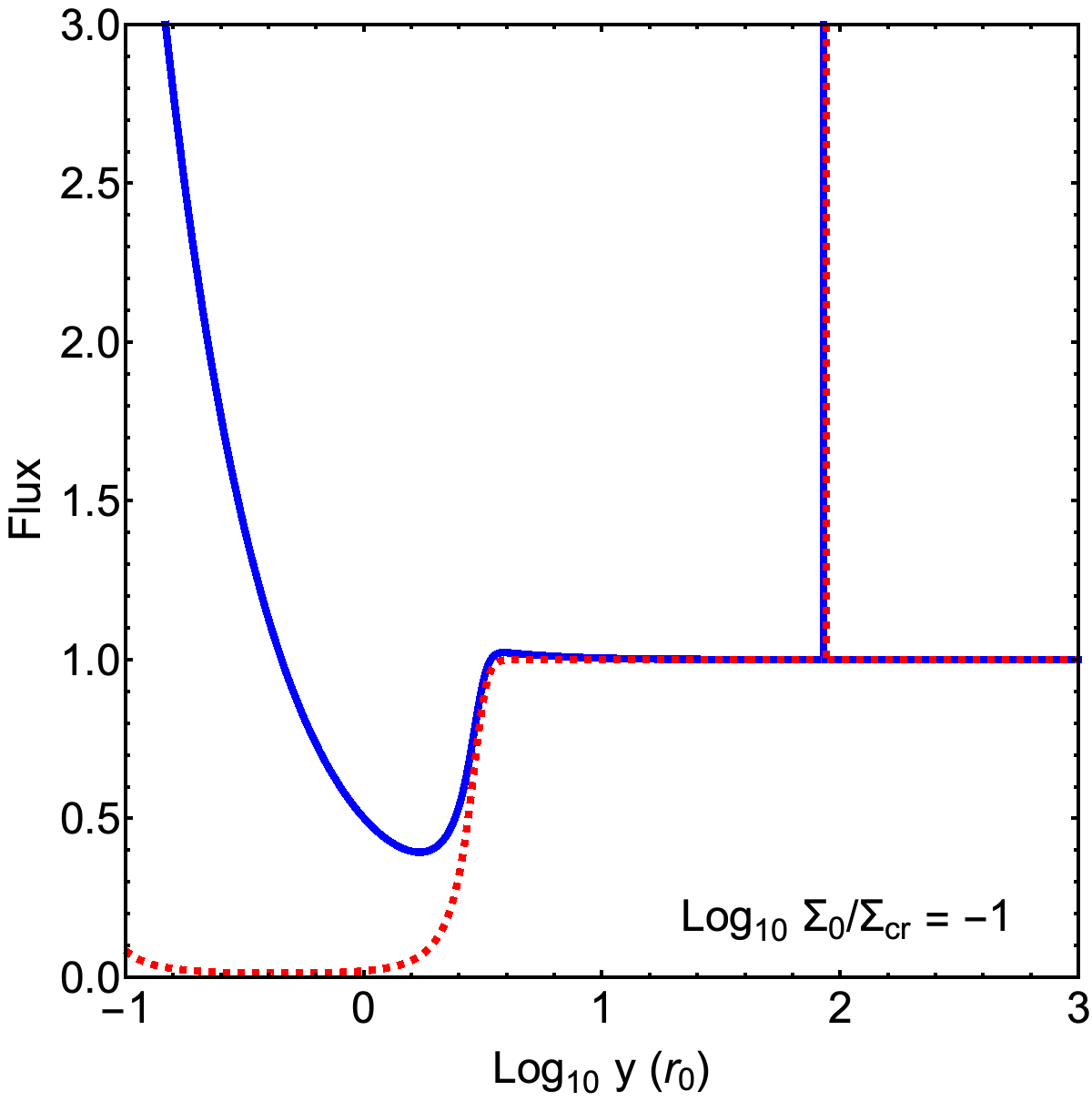}}\\
\centering
{\includegraphics[width=0.3\textwidth]{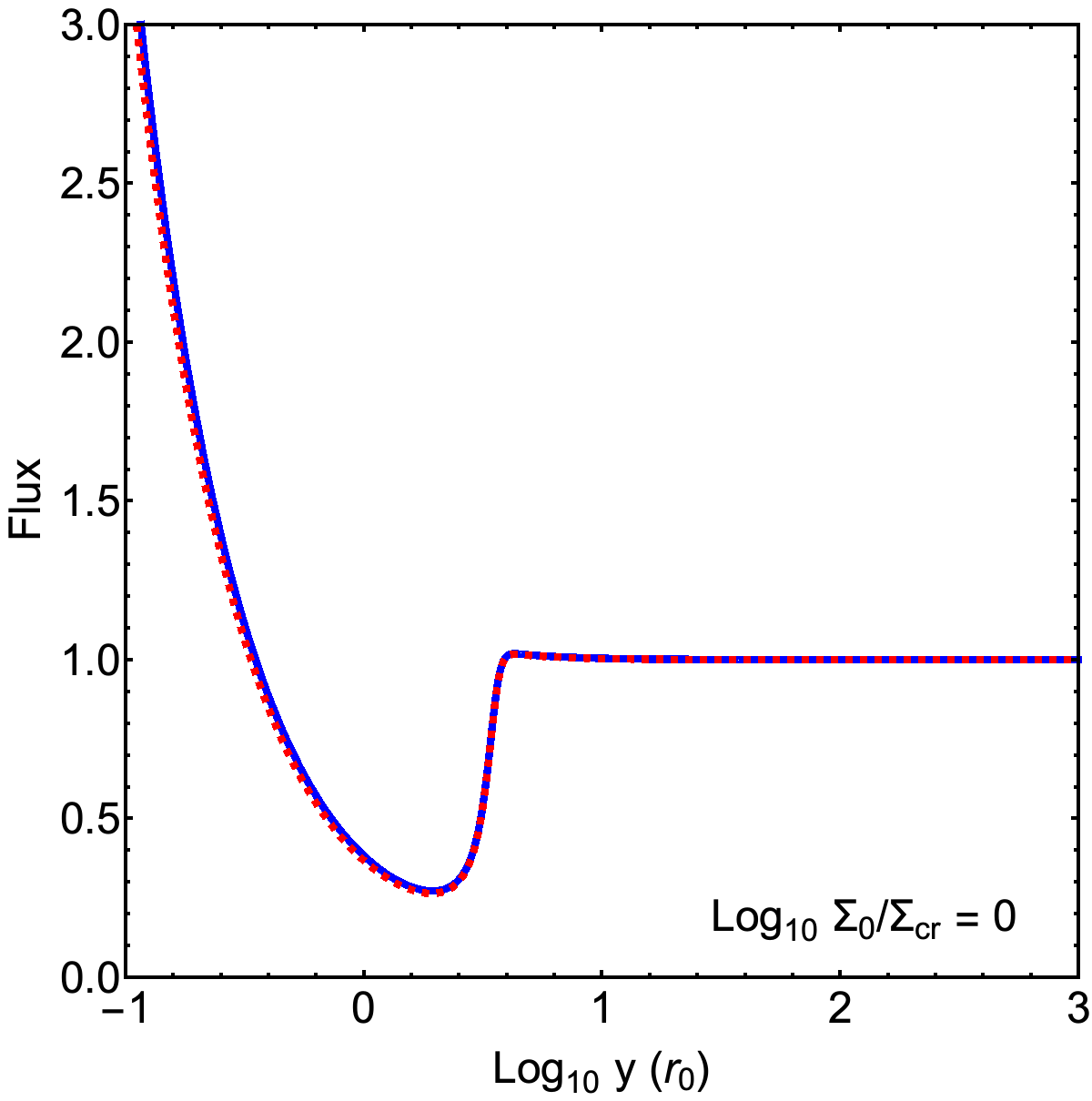}
\includegraphics[width=0.3\textwidth]{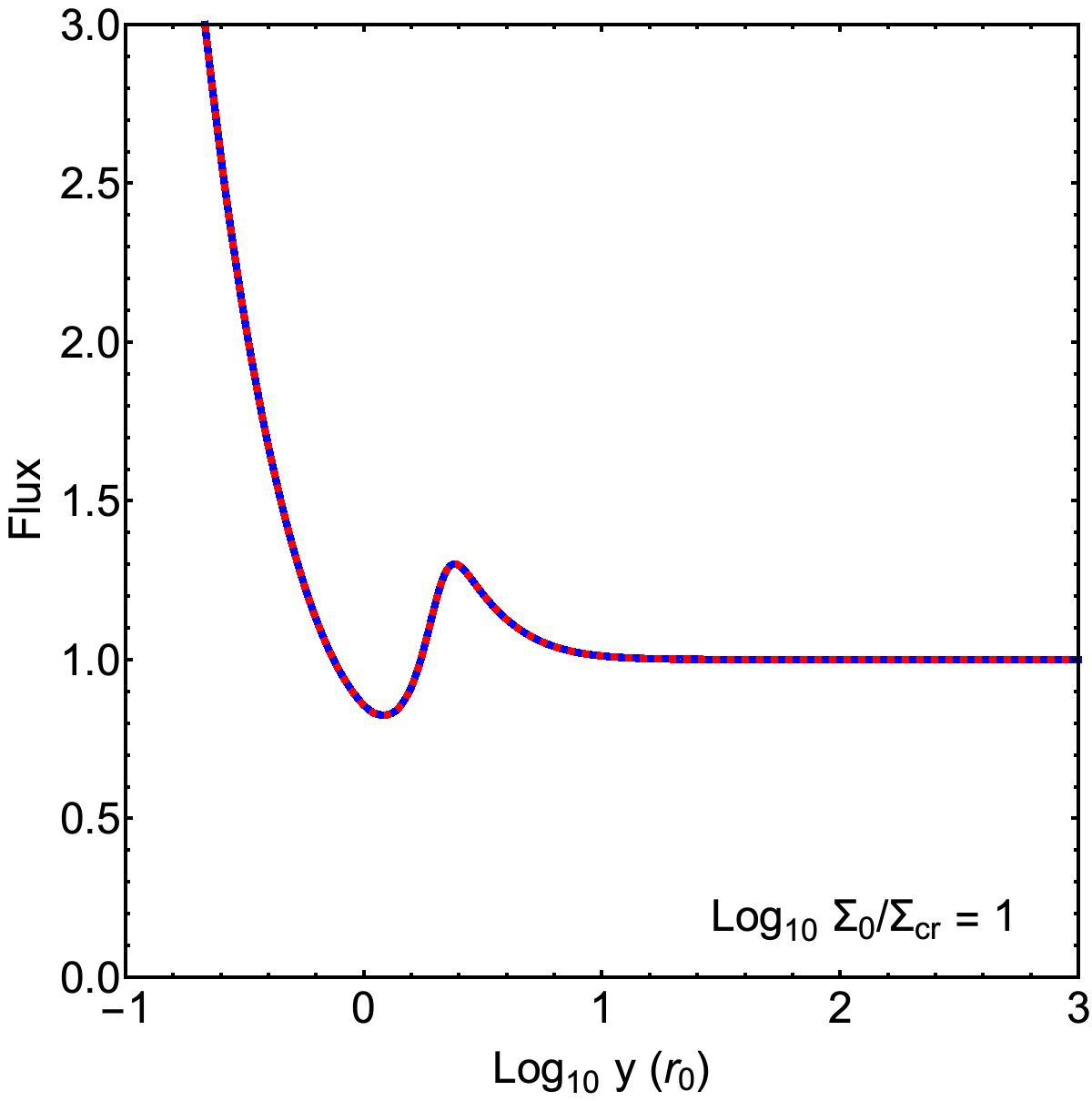}
\includegraphics[width=0.3\textwidth]{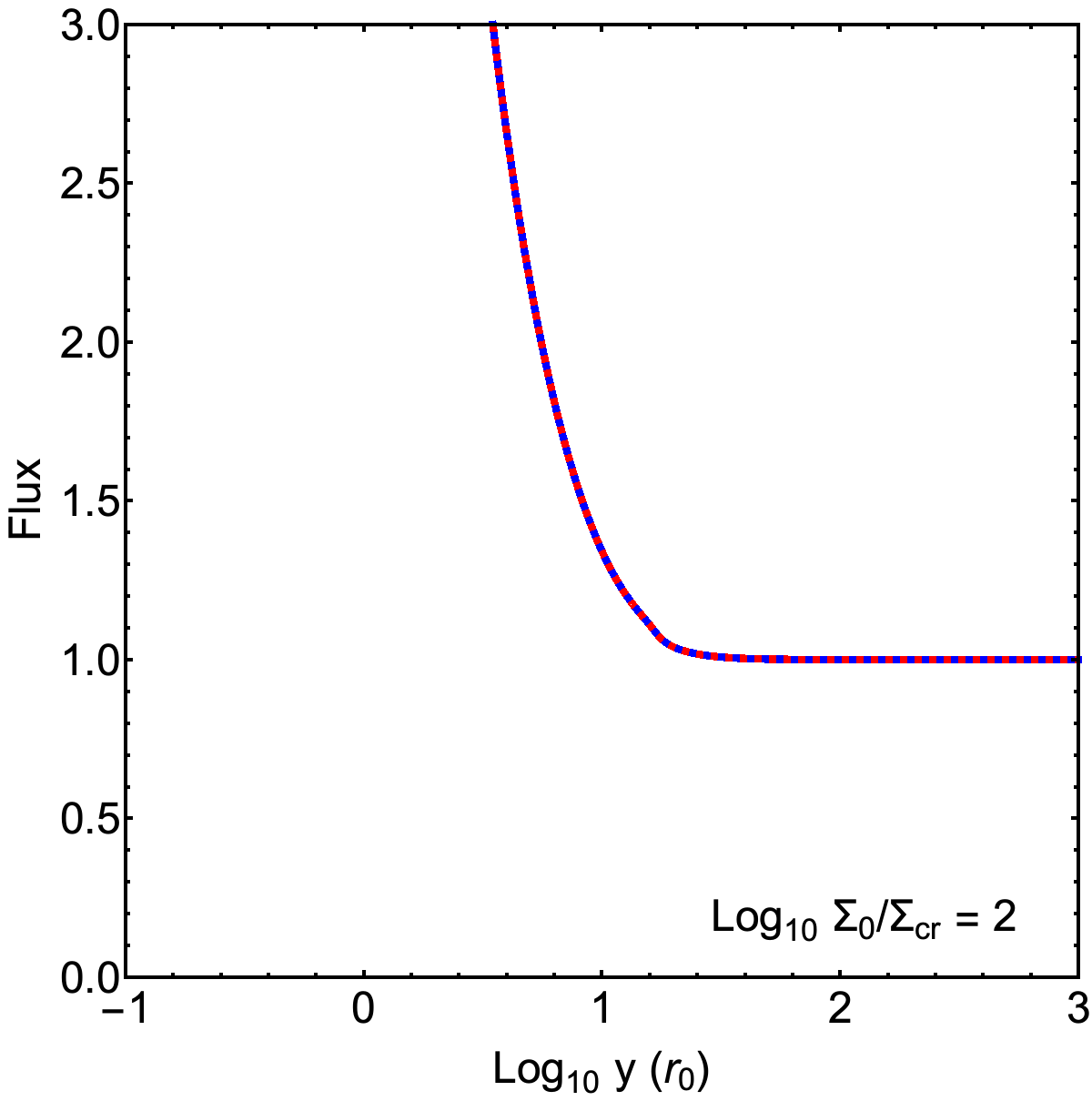}}
\caption{The flux of a point-like source, normalised to the flux in the absence of an occulting cloud, as a function of (the Log of the) source position, $y$, from equation (19), expressed in units of $r_0$. In all cases the cloud mass is $10^{-4}M_*$, but with different values of the central column-density as indicated in the lower right-hand corner of each panel.  The solid, blue curve is appropriate to a completely transparent cloud, and the dashed, red curve is for a cloud with a central optical depth to extinction of $1{,}000$. We note that the vertical line at $y\simeq 100\,r_0$ in the top-right panel is a real feature -- it is the radial caustic of the lens mapping -- but it is very narrow and would appear washed out in the case of non point-like sources.}
\label{figure:fluxprofiles}
\medskip
\end{figure*}

Analogous to figure 1, these estimates allow us to delineate domains in the mass-radius plane where the clouds can be classified as: weak lenses; strong gas lenses; or, strong gravitational lenses. Figure 2 shows that classification for $n=3/2$ polytropes, for both Galactic and cosmologically-distant clouds, overlaid with the theoretical gas cloud models shown in figure 4 of \citet{walkerwardle2019}.

Also shown in figure 2 are the boundaries between opaque and transparent clouds, for the smallest and largest material transparencies decided on in \S2.1.2. The effect of these boundaries must be considered in combination with the nature of the lensing, in order to anticipate the character of the light-curves that would result from an occultation, as follows. In the case of strong gravitational lensing the dominant images are formed beyond the limb of the cloud, where the optical depth is zero regardless of the cloud's material opacity, and extinction can have only a minor influence on the light-curves.

If the cloud is not a strong gravitational lens then images will form inside the limb and will thus be subject to some extinction; but it need not be a large effect, even if much of the cloud is opaque. The key point to notice is that if the gas lensing is sufficiently strong then during an occultation the dominant images will form close to the limb of the cloud --- because the refraction angle is zero at the limb, but increases rapidly as one moves inwards, so rays can only reach the observer if they pass near the limb. And near the limb the column of gas and dust is relatively low. Depending on the details -- i.e. the material opacity, the column-density profile, and the strength of the gas lensing -- the dominant images could then be subject to either high- or low-levels of extinction. Thus we see that the material opacity is guaranteed to have a major effect on the light-curve only if both gaseous and gravitational refraction are weak enough that the dominant images traverse the central, opaque region of the cloud. In that case, of course, occultations by opaque clouds will generate extinction events.

With those points in mind, by examining figure 2 we arrive at the following conclusions. If the material transparency is low then all of these models are totally opaque in V-band; and even if the material transparency is high then only the larger clouds are transparent. Given that our ``high transparency'' case corresponds to the absolute maximum that is physically possible (i.e. only the molecular contribution to the opacity), at the low mass end of the distribution the \citet{walkerwardle2019} models are guaranteed to be opaque. Naturally the lensing characteristics differ greatly between Galactic and cosmological contexts. In the Galactic case the larger clouds are seen to form only weak lenses, so depending on the material transparency those clouds could result in occultations that can be classified as pure extinction events, or occultations with very little effect on the measured source flux. By contrast small clouds within the Galaxy are seen to act as strong gas lenses, so that significant magnification (and also demagnification) is expected --- as per \citet{draine1998,2001ApJ...547..207R}. However, as already noted, these models must also be opaque and the resulting light-curves should manifest both extinction and gas lensing. Even in the Galactic context, gravitational lensing starts to become important for the smallest of the \citet{walkerwardle2019} model clouds, pushing the dominant image beyond the limb of the cloud so that extinction plays a minor role in determining the appearance of occultations. Figure 2 also demonstrates that in the cosmological context essentially all of the models from figure 4 of \citet{walkerwardle2019} are very strong gravitational lenses and thus would yield light-curves very similar to those of a point-mass gravitational lens \citep[e.g.][]{1986ApJ...304....1P}.

\subsubsection{Example flux maps for gaussian lenses}
In this paper we utilise a model with a gaussian column-density profile -- because the simulations presented in the next section require rapid solution of the lens equation -- which does not have a finite extent (radius). But it is straightforward to classify the regimes of lensing and extinction behaviour based on any two of the parameters $\{ M, \Sigma_0, r_0\}$. In the previous section (\S2.2) we used $\{\Sigma_0, r_0\}$ (equations 20) to classify the lensing behaviour, but in \S3 it is more convenient to use $\{M, \Sigma_0\}$ instead. In terms of those parameters the criterion for a strong gravitational lens remains the same, namely $\Sigma_0>\Sigma_{cr}$, whereas the criterion for a strong gas lens becomes $M/M_*<2(\Sigma_0/\Sigma_{cr})^2$.

Any given cloud occulting a source can generate a variety of light-curves, depending on the impact parameter of the event ($b$) and the effective transverse velocity of the cloud relative to the source ($v_\perp$). For axisymmetric cloud models the effect of these kinematic parameters can be readily imagined because they describe the (projected) radial position of the source, $y$, as a function of time, $t$, via $y^2=b^2+v_\perp^2 t^2$. Given the appropriate flux map for the cloud -- i.e. the source flux as a function of $y$ -- the entire family of possible light-curves can be anticipated. We note that in the absence of any extinction the flux map could alternatively be called the ``magnification map'', which terminology is familiar in the context of gravitational lensing.

Figure \ref{figure:fluxprofiles} illustrates a variety of flux maps for point-like sources seen through a lens with a gaussian profile, depending on the strength of the gas lensing, gravitational lensing and extinction. Modulo the logarithmic scaling of the horizontal axes, these plots can be thought of as half of the light-curve (the egress) for events with zero impact parameter. The cloud mass is $10^{-4}M_*$ in all cases, with the central column varying by a factor of 10 between consecutive panels from $10^{-3}\Sigma_{cr}$ (top left) to $10^{2}\Sigma_{cr}$ (bottom right). And in each panel the map is shown for both $\kappa_\lambda\rightarrow0$ (transparent case; blue curve) and a lens with $\kappa_\lambda^{-1}=10^{-3}\Sigma_0$ (opaque case; red, dashed curve). For the lowest column-density shown the lensing is weak and thus in the transparent case there is almost no discernible effect on the source flux; in the opaque case occultations yield pure extinction events. At a column of $10^{-2}\Sigma_{cr}$ the cloud is already a strong gas lens, manifesting a radial caustic at $y\simeq 0.3\, r_0$ in the transparent case, whereas in the opaque case the map remains close to that of the opaque, weak lens. The region of demagnification at $y\sim r_0$ is noteworthy: gas lensing can produce demagnification, but gravitational lensing cannot. Increasing the column further to $0.1\,\Sigma_{cr}$ makes for a very strong gas lens with the radial caustic pushed out to $y\simeq 10^2\, r_0$; the map for the opaque case is still dominated by extinction, but the tangential caustic has a non-negligible influence near the origin. When the central column reaches $\Sigma_{cr}$ the cloud is just at the transition to strong gravitational lensing, while the gas lensing is very strong indeed and is the dominant influence on the shape of the flux map. At this point the gas lensing is so strong that the dominant images are formed at large radii; there even the opaque cloud imposes very little extinction and so the flux map of the opaque cloud is very similar to that of the transparent cloud. As the central column increases further the differences between the opaque and transparent cases become too small to discern with the eye, as the dominant images form at still larger radii. At a central column of $10\,\Sigma_{cr}$ the flux map is that of a hybrid gas and gravitational lens; the net demagnification is now limited to modest values because it is offset by the magnification due to the gravitational refraction. For the largest central column shown in figure 3 ($100\,\Sigma_{cr}$) the demagnification region has completely vanished and the flux map is very close to that of a point mass gravitational lens: the largest differences relative to a point mass are at the level of a few percent, manifest at $y\simeq 20\, r_0$.

\begin{figure*}
\includegraphics[width=2 \columnwidth]{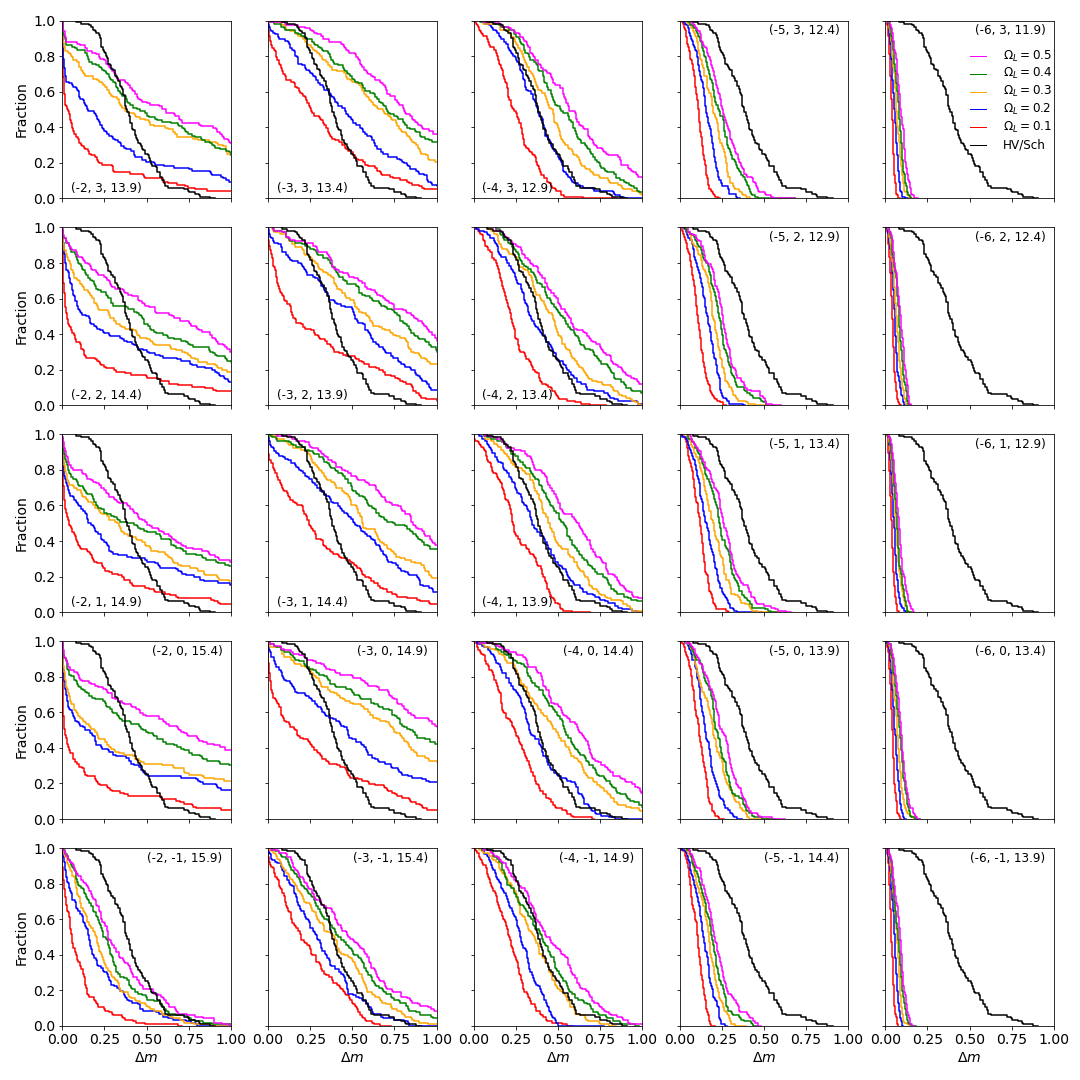}
\caption{Fraction of the synthetic light curves changing by a given magnitude threshold for a selection of cloud models with gaussian column-density profiles, characterised by the parameters $(\log_{10}\,M, \log_{10}\,\Sigma_0, \log_{10}\,r_0)$ as indicated in each panel. We note that $M$ and $\Sigma_0$ define the model, and an approximate value of $\log_{10}\,r_0 ({\rm cm})$ is quoted in addition, for the convenience of readers. The layout of the grid is as follows: columns correspond to clouds of fixed mass, with $\log_{10}\,M({\rm M_\odot}) = \{-2, -3, -4, -5, -6\}$, left to right; rows correspond to clouds of fixed central column-density, with $\log_{10}\,\Sigma_0({\rm g\,cm^{-2}}) = \{3,2,1,0,-1\}$, top to bottom. Lines of different colours represent different values of the cosmological density $\Omega_l$ that the clouds make up, as per the legend in the top-right panel. We note that the two largest values, $\Omega_l=0.4, 0.5$ are actually greater than the total matter content of the background cosmological model that we are using, and cannot be physically realised. For comparison, the distribution in the \citet{1993MNRAS.260..202H} sample, as read off Figure~5 in \citet{1993A&A...279....1S} is shown in black in each panel. This figure is appropriate to clouds that are as transparent as possible, with a material transparency of $\kappa_\lambda^{-1}=768\,\mathrm{g}\,\mathrm{cm}^{-2}$ (see \S2.1.2).}
    \label{figure:surplotsh2}
\end{figure*}

All of the maps shown in figure 3 are for point-like sources and if we consider extended sources then there is substantial additional variety in the resulting flux maps. However, the effect of a non-zero source size is simply to convolve the (two-dimensional) point-source flux map with the intensity profile of the source, so the influence can be readily conceived as a simple smearing of the point-source flux map.  For sufficiently large sources the effects of any one lens can be made arbitrarily small, but if there are many clouds occulting a source then it is their aggregate effect that is of interest --- e.g. the total extinction, as discussed in \S4.2.

\section{Cosmological nanolensing of quasars}
\label{section:application}
\citet{1993Natur.366..242H} suggested that gravitational microlensing by a cosmological population of black holes could explain the observed optical variability of quasars. That idea was explored theoretically by \cite{1993A&A...279....1S}, using simulations of microlensing by point-mass objects, with source redshifts and temporal sampling matched to the data underlying \cite{1993MNRAS.260..202H}. By comparing the variability amplitudes (i.e. full variation from  minimum to maximum) in simulated light curves against the data, \cite{1993A&A...279....1S} was able to rule out a large cosmological population of point-like lenses with masses in the range $10^{-3}\,\mathrm{M}_\odot\,\la\, M\, \la\, 3\times 10^{-2}\,\mathrm{M}_\odot$, under certain assumptions about the underlying cosmological model and intrinsic quasar properties.

\begin{figure*}
\includegraphics[width=2 \columnwidth]{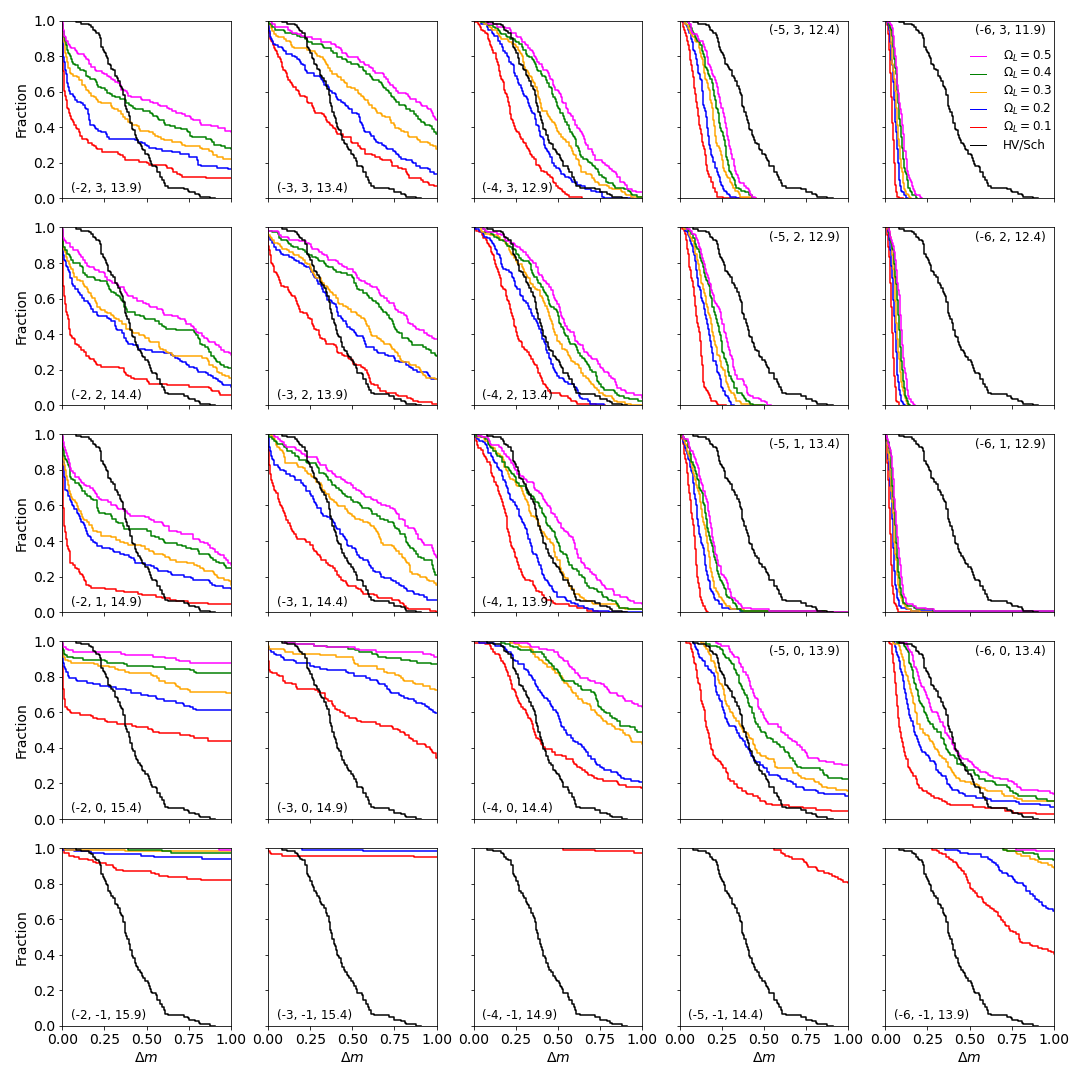}     \caption{Same as Figure~\ref{figure:surplots}, but with low material transparency,  $\kappa_\lambda^{-1}\approx 3.6\times10^{-3}\,\mathrm{g}\,\mathrm{cm}^{-2}$, as would be appropriate for the local, Galactic ISM.}
    \label{figure:surplots}
\end{figure*}

In this section we repeat the analysis of \cite{1993A&A...279....1S}, but for the case of non-point-like lenses. Specifically: we consider compact gas clouds, and we include the effects of gas lensing and extinction. As our background geometry we adopt the Planck cosmological model \citep{2016A&A...594A..13P} for the computation of angular-diameter distances and volume elements. However, in other respects we follow in detail the approach described by \cite{1993A&A...279....1S}; in particular we adopt the magnification-multiplication approximation when calculating the total microlensing magnification due to multiple lenses along the line of sight. Although inadequate to describe gravitational lensing in the case of near-unit optical depth, it is nevertheless suitable for computing the statistics of significant changes in the magnification on short time scales -- years, say -- which are dominated by changes in the relative position with respect to the lens closest to the line of sight. The multiplication approximation is a useful and physically motivated method to interpolate from one such lens to another. 

In our simulations, clouds are distributed at random with a specified, uniform mean comoving number-density. We compute the magnification factor numerically for clouds that are closer than either 15 length scales ($r_0$) or 15 Einstein radii (whichever is the greater) to the nearest edge of the source. For all other clouds the magnification is very close to unity and very little error is incurred by utilising the analytic result for a point mass lens; we therefore utilise that approximation for these clouds. For practical reasons noted in~\S\ref{section:phenomenology} we model the source as a disc of uniform brightness and assume a radius of $R_s=3\times 10^{15}\,\mathrm{cm}$, which is three times larger than used by \cite{1993A&A...279....1S}; this choice is discussed in Appendix A.

To generate light-curves, the sources are all assumed to move at $400\,\mathrm{km}\,\mathrm{s}^{-1}$ and lenses are assigned random velocities drawn from a gaussian distribution with an r.m.s. of $400\,\mathrm{km}\,\mathrm{s}^{-1}$ per component; the observer is stationary. The vast majority of our synthetic light curves resemble a series of Paczy\'nski curves \citep[][]{1986ApJ...304....1P}, similar to point-mass gravitational lenses, and a minority of light curves manifest extinction events. Following \cite{1993A&A...279....1S}, we compare our simulated light curves to the data through the variability amplitude distribution, having matched the redshift distribution and temporal sampling of our simulations to those of the \cite{1993MNRAS.260..202H} sample. We include all sources, regardless of the average or minimum magnification in their light curve, when deriving the amplitude distribution, which corresponds to the 'volume-limited' case (as opposed to 'flux-limited' one) in the calculations undertaken by \cite{1993A&A...279....1S}.

Figures~\ref{figure:surplotsh2} and~\ref{figure:surplots} present the fraction of light curves with variability amplitude above a given threshold, for a variety of cloud models. Each panel in these figures corresponds to a particular combination of cloud mass and central column-density, with the various curves within each panel showing results for different values of the assumed cosmological density in clouds of that type, $\Omega_l$. Unsurprisingly, figures \ref{figure:surplotsh2} and \ref{figure:surplots} show that for each cloud model the variability amplitudes increase systematically with $\Omega_l$. We note that our adopted cosmology has a total matter content that is less than the two largest values ($\Omega_l=0.4, 0.5$) used for the simulations shown in these figures, so the corresponding universes cannot be physically realised. Figure~\ref{figure:surplots} is appropriate to clouds having a transparency similar to that of the local Galactic ISM \citep{bohlin1978}, while figure~\ref{figure:surplotsh2} is appropriate to much more transparent gas, where the material transparency is set at the upper limit set by molecular absorption (as discussed in \S2.1.2). 

Figure~\ref{figure:surplots} demonstrates a strong change in behaviour, at fixed cloud mass, as the central column-density, $\Sigma_0$, decreases. For high central columns the behaviour is essentially that of point-mass gravitational lenses, while for very low central columns the variability is dominated by extinction and exhibits very large amplitudes for clouds of all masses.

Much like~\cite{1993A&A...279....1S}, our results readily lend themselves to excluding substantial populations of certain types of lenses. Indeed, as noted above, for the majority of our models -- those with sufficiently high central columns $\Sigma_0\gg \Sigma_{cr}$ -- the effect of gas clouds is much the same as that of point-like lenses, and consequently our constraints are broadly similar to those of \cite{1993A&A...279....1S, 2003A&A...399...23Z, 2003A&A...408...17Z}. In other words: a cosmologically significant population of dense gas clouds with masses $M \gg 10^{-4}\,{\rm M}_\odot$ can be excluded because such a population would result in variation amplitudes that are much larger than observed. This incompatibility arises because high-mass gravitational lenses have Einstein radii that are larger than the sources, so high magnifications commonly occur.  

Also in conflict with the data are low-transparency gas clouds with low central column-densities, $\Sigma_0\la 1\;{\rm g\,cm^{-2}}$, for clouds of all masses within the range investigated (figure~\ref{figure:surplots}). For these clouds the very large variation amplitudes seen in our simulations are primarily due to extinction. We caution that these large extinction fluctuations are in part attributable to the very extended (infinite!) radius of the gaussian column-density profile that we are using as a model; and that aspect of the model is, of course, unphysical.

Less objectionable are the very low mass clouds, $M \ll 10^{-5}\,{\rm M}_\odot$, which (excepting the low column-density, low transparency clouds noted above) all produce little variability. These low-mass lenses have Einstein ring radii that are much smaller than the source radii, so significant magnification does not occur. We cannot exclude a high abundance of such clouds. But on the other hand there doesn't seem to be much motivation to consider such a population, given that they cannot explain the observed quasar variability. 

At the interface between the benign-but-uninteresting low-mass clouds and the high-mass clouds, which produce too much variability, figures~\ref{figure:surplotsh2} and ~\ref{figure:surplots} show that there are some combinations of cloud mass, column-density and abundance for which our simulations produce variability amplitude distributions that are similar to the observed quasars. These parameter combinations are ``preferred'' in the sense that they have the potential to explain the optical variability of quasars.

In determining the parameters of our preferred models it is important to bear in mind that the observed distribution is not corrected for measurement uncertainties. Nor are those uncertainties included in our simulations, because the information provided with the data is not sufficient to allow us to do that. Crudely speaking the effect of measurement errors of $\sigma$ magnitudes is to put a floor of $\sim 2\sigma$ on the measured variability amplitude of all sources, and this will shift the observed probability distribution by the same amount. Examining the black curve that is plotted in each panel of figures 7 and 8 suggests that the photometric errors in the data might be as much as $\sigma\sim0.05\;{\rm mag}$ --- a figure that is consistent with the estimates in \citet{1996MNRAS.278..787H}. Making allowance for that possibility, our simulations would best match the data if $M\la 10^{-4}\,{\rm M}_\odot$ (with $\Omega_l\ga 0.1$) and $M\ga 10^{-5}\,{\rm M}_\odot$ (with $\Omega_l\la 1$). This preferred range applies to all of the central column-densities we investigated in the case of high transparency gas clouds (figure~\ref{figure:surplotsh2}), and for central columns $\Sigma_0\ga 10\;{\rm g\,cm^{-2}}$ in the case of low transparency material (figure~\ref{figure:surplots}).

\section{Extinction due to dense clouds}
\label{section:grey}
The previous section confined attention to the variability introduced by a population of dense gas clouds, but there is also an effect on the average flux of distant sources --- even for dense clouds with Einstein radii in excess of their physical size. Parts of the image plane with strong extinction inevitably map onto some areas of the source plane, regardless of the effects of gravitational and gas lensing, and therefore the total flux is lower than it would have been if the lenses were completely transparent.\footnote{This effect is mentioned in footnote 2 of \cite{1999ApJ...519L...1M}.}

\subsection{The reddening curve}
Reddening is a key signature of dust extinction. But for sufficiently dense and compact gas clouds, the extinction that is introduced is very nearly grey even when the specific opacity of the material is strongly wavelength dependent. The reason is that transmission through the central part of a highly opaque lens is effectively zero over a broad range of wavelengths. In this case the position of the outer edge of the opaque region changes with wavelength, but when most of the lens is opaque there is little room for this change.

\begin{figure}
\includegraphics[width=\columnwidth]{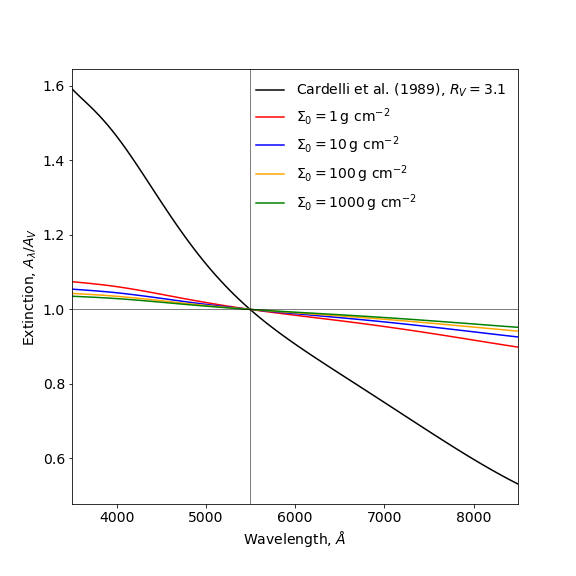}
\caption{The reddening curve (black line) for diffuse gas with a total-to-selective extinction ratio $R_V=3.1$, as appropriate to the local, diffuse ISM  \citep{1989ApJ...345..245C}. Also shown are the reddening curves for high-column density gas clouds, with gaussian profiles and various values of $\Sigma_0$ (as per the legend), \emph{having the same material opacity as the local, diffuse ISM.}}
    \label{figure:reddening}
\end{figure}

We can define a wavelength-dependent opaque radius where the optical depth through the lens is unity $r_\lambda\,:\,\Sigma(r_\lambda)=\kappa_\lambda^{-1}$, which, for the gaussian lens of eq.~(\ref{gaussianlens}) with $\Sigma_0>\kappa_\lambda^{-1}$, is
\begin{eqnarray}
r_\lambda=r_0\sqrt{\log \kappa_\lambda\Sigma_0}.
\end{eqnarray}
The (image plane -- please see below) effective optical depth to extinction of a \emph{population} of such lenses with number density $n$ and line-of-sight depth $D$ is
\begin{eqnarray}\label{taupop}
\tau^{pop}_\lambda=n D\int\md r\, 2\pi r \left[1-\mathrm{e}^{-\kappa_\lambda\Sigma(r)}\right]\approx \pi n D r_\lambda^2\\=\pi n D r_0^2\log\kappa_\lambda\Sigma_0\,\hspace{1.7cm}\mathrm{(gaussian)}\label{taupopgaussian}
\end{eqnarray}
where the approximation replaces the $[1-e^{-\tau(r)}]$ factor with a step function at position $r_\lambda$ as appropriate if $\tau$ increases rapidly around $\tau=1$. In fact, for a gaussian lens the integral can be calculated exactly with $\log\kappa_\lambda\Sigma_0$ acquiring a correction of $\tilde\gamma-\mathrm{Ei}(-\kappa_\lambda\Sigma_0)$. Either way this is a very slow function of $\kappa_\lambda$ -- and, thereby, $\lambda$ -- when $\Sigma_0$ is well above the material transparency $\kappa_\lambda^{-1}$. Figure~\ref{figure:reddening} compares the familiar reddening curve for the local, diffuse Galactic ISM \citep{1989ApJ...345..245C} with that of compact gas clouds \emph{of the same composition}, for various assumed central column-densities; it is evident that the reddening effect of the dust is greatly reduced.

For completeness, in the cosmological case with a constant co-moving number density of clouds $n_\mathrm{comov}=\Omega_l \rho_\mathrm{cr}/M$, the surface density $nD$ would need to be replaced by an integral
\begin{eqnarray}
n D\to N(z_s)=n_\mathrm{comov} D_\tau(z_s)=\frac{3H_0^2\Omega_l}{8\pi G M} D_\tau(z_s)\\
\mathrm{~with~~}D_\tau(z_s)\equiv\int\limits_0^{z_s}\frac{\md^3 V_\mathrm{comov}(z)}{\md z\, \md^2\Omega}\frac{\md z}{D_A^2(z)},
\end{eqnarray}
e.g. employing the $\md^3 V_\mathrm{comov}(z)/\md z\, \md^2\Omega$ volume as conveniently realised in the {\tt Python\/} programming language with the {\tt differential\_comoving\_volume()\/} method of the {\tt astropy.cosmology.FLRW\/} class \citep{astropy:2018}.

An important qualitative point to note is that the gaussian column-density profile used to construct the curves in figure \ref{figure:reddening} extends to infinite radius. More realistic cloud models \citep[e.g.][]{walkerwardle2019} have a finite extent, and with sufficiently high central optical depths such models may exhibit even less reddening than shown here.

\subsection{Extinction estimate for large sources}
The optical depth defined on the image plane, $\tau^{pop}$, is an appropriate measure of extinction when gravitational and gas lensing is weak but should be corrected (which we will mark by a hat, $\hat\tau^{pop}$) to account for the relative areas of differently extinguished portions of the image plane when mapped onto the source plane. In the case of the gas clouds considered in this paper this mapping is correlated with the degree of extinction, making correction non-trivial. Here it should be reiterated that the individual images have a very small angular separation and would be correspondingly difficult to resolve; in practice, then, only the total source flux is measured. 

In the case of a sufficiently large source, with radius $R_s\gg\,R_E\,(\Sigma_{cr}/\Sigma_{pop})^{1/2}$, we expect there to be many lenses projected onto the source, resulting in a net magnification (relative to an empty beam) of $\langle\mu\rangle\simeq(1-\Sigma_{pop}/\Sigma_{cr})^{-2}$, as would have been expected of a smooth matter distribution. The only difference with the latter is that there would be a dark spot at the location of each of the opaque clouds, and therefore the source plane effective optical depth to extinction is equal to its image plane, geometric value~(\ref{taupop}, \ref{taupopgaussian}),
\begin{eqnarray}
\hat\tau^{pop}_\lambda\approx\tau^{pop}_\lambda\approx\left\langle\frac{\Sigma_{pop}}{\Sigma_{cr}}\frac{r_\lambda^2}{R_E^2}\right\rangle.
\end{eqnarray}
If the cloud abundance is sufficiently large the optical depth to extinction may be non-negligible.

\subsection{Supernova dimming}
\label{subsection:SNIadimming}
The foregoing points are relevant to the interpretation of ``standard candles'' at cosmological distances, such as type Ia supernovae \citep[e.g.][]{1998AJ....116.1009R,1998Natur.391...51P}. The low levels of reddening that are observed in distant SNe Ia \citep[e.g.][]{1999ApJ...525..583A,2002A&A...384....1G,2004ApJ...607..665R} provide tight constraints on the extinction contribution from dust associated with intervening diffuse gas if it is similar to the local, Galactic ISM. In the case of dense gas clouds, however, we have shown that the extinction is very nearly grey even if the clouds are made of material similar to the local ISM (figure \ref{figure:reddening}), so that low levels of reddening do not exclude high levels of extinction. That raises the question of whether dense gas clouds might actually be having a significant effect on the shape of the magnitude-redshift relation for type Ia SNe? In this section we address that question.

Attempts to place constraints on the extinction contributed by dense clouds are hampered by a lack of knowledge about their properties. Some simple physical models of the internal structure of dense, Galactic clouds have recently been constructed \citep[][]{walkerwardle2019}, but as yet there are no reliable models for the formation and evolution of such objects. Some hypothetical evolutionary scenarios can be constrained using the supernovae photometry. For example: clouds which form at high redshift and do not subsequently evolve would yield a redshift-dependent extinction much like that of a smooth background distribution of dust with constant comoving density \citep[c.f.][]{2002A&A...384....1G}. In models of this type the extinction increases rapidly with redshift, because the density scales as $(1+z)^3$, and would be easy to recognise; such models are excluded by the available data on SNe Ia \citep[][]{2004ApJ...607..665R}. However, this very simple scenario does not offer a plausible description of the evolution of the dense clouds under consideration at $z\ga 1$. We note, in particular, that at $z\ga 1$ the microwave background radiation will play an increasingly important r\^ole in the thermodynamics of the outer, cooler regions of the cloud, where nucleation and precipitation of hydrogen dust (i.e. particles of solid H$_2$) takes place \citep[][]{walkerwardle2019}.

\subsubsection{Extinction fluctuations}
One constraint that any acceptable model should satisfy is that the \emph{variations} in extinction, across different lines of sight at the same redshift, must be small, because of the small photometric scatter around the mean that is observed for type Ia SNe. This requirement parallels the situation with gravitational lensing of type Ia SNe: the small observed dispersion in fluxes places an upper limit on the mass of the elementary lumps of matter, such that the luminosity distance is well approximated by that of a homogeneous universe \citep[][]{1999ApJ...519L...1M,2018PhRvL.121n1101Z,2020OJAp....3E...1H}. The latter condition is satisfied if we expect to have a number $\langle N\rangle\gg 1$ elementary lumps of matter within the ``beam'' --- i.e. inside the cone defined by the observer, at the vertex, and the photosphere of the source as the base. This is commonly referred to as the ``filled beam'' case, in contrast to the ``empty beam'' circumstance $\langle N\rangle\ll 1$ that is expected when the elementary lumps of matter are sufficiently massive. In the context of gravitational lensing the mass within the beam translates directly into beam convergence, $\varkappa$ (see \S2), and in the filled beam case we expect small fractional fluctuations in the beam convergence
\begin{equation}
{{\delta \varkappa}\over{\varkappa}}\sim {1\over{\sqrt{\langle N\rangle}}} \ll 1.
\end{equation}
The correspondence to extinction fluctuations is now obvious, because the expected optical depth to extinction is $\sim\langle N\rangle \times r_\lambda^2/R_s^2$, and the fluctuations around the mean are smaller by a factor $\sqrt{\langle N\rangle}$. We note that the increased extinction caused by having $\sim \sqrt{\langle N\rangle}$ additional clouds in the beam (relative to the mean) is offset by the additional beam convergence (hence magnification) that they introduce --- i.e. the photometric fluctuations of the two effects are totally anticorrelated. 

The cloud mass and abundance, together with the assumed background geometry, determine the expected number of clouds in the beam. Whatever background is assumed, a useful reference point can be established by placing the entire matter content in the form of dense clouds of a single mass; for the (Planck) cosmology that we adopted in \S3 that circumstance corresponds to $\langle N\rangle = 1$ at a cloud mass of $\sim 10^{-4}\,{\rm M}_\odot$ (for $R_s=3\times 10^{15}\,{\rm cm}$ at $z=1$).

 \subsubsection{Can extinction be a large effect?}
We have already given general expressions for the amount of extinction that is introduced by a population of dense clouds. There is, however, a specific question that is of particular interest in connection with the magnitude-redshift relation of type Ia SNe: can the extinction be large enough to reconcile the observations with a non-accelerating universe? In this section we address that question. 

The observed dimming amounts to a difference, relative to a flat, matter dominated universe, of about $0.5$ magnitudes at $z\simeq1$. To produce this much extinction requires each cloud to obscure quite a large area. And the opaque radius cannot be larger than the cloud radius, so in order to produce a dimming as large as is observed we require a cloud population that has a covering fraction of at least $0.5$ at $z=1$. In the case of a flat universe dominated by a population of dense clouds that translates directly to a requirement that the individual clouds have an average column-density
\begin{equation}
{{M}\over{\pi R^2}}\;\la\; 0.2\, {\rm g\,cm^{-2}}.\label{eq:minimumcolumn}
\end{equation}
We emphasise that this is a limit on the average column-density of a single cloud -- the peak (central) column-density of a cloud can be much larger -- so it is not a region of parameter space that is excluded by the results of \S3 (figure \ref{figure:surplots} in particular). It is in fact difficult to relate the constraint on mean column (\ref{eq:minimumcolumn}) to a gaussian cloud model, because that model extends to infinite radius.

For the convenience of readers we translate our column-density limit (\ref{eq:minimumcolumn}) into a lower limit on the cloud radius of
\begin{equation}
R\;\ga\; 6\,\left({{M}\over{10^{-5}\,{\rm M_\odot}}}\right)^{1/2}\; {\rm AU},
\end{equation}
where we have scaled to the low-mass end of the preferred range identified in \S3 as being able to reproduce the observed, optical variability of quasars.

We are not aware of any fundamental reason why dense gas clouds should not have radii greater than the limit just derived, so it is in principle possible that extinction by dense clouds could have a substantial influence on the observed magnitude-redshift relation of type Ia SNe. There are, however, two caveats to add. First, the limit given in equation~(\ref{eq:minimumcolumn}) is orders of magnitude smaller than the characteristic value ($140\;{\rm g\,cm^{-2}}$) estimated for clouds within galaxies by \citet{1999MNRAS.308..551W}, from consideration of the destructive collision rate between clouds. It might be possible to reconcile these two very different estimates for the mean column if the mass of each cloud is strongly concentrated towards the centre. In that case complete destruction would require a physical collision with an impact parameter that is small compared to the cloud radius, leading to a much larger estimate of the mean column. The second caveat is that for $M\sim 10^{-5}\,{\rm M_\odot}$ our lower limit on cloud radius is only slightly smaller than the largest model of that mass obtained in theoretical structural modelling \citep[figure 5 of][]{walkerwardle2019}.

\subsubsection{Other constraints on supernova dimming by dense clouds}
The possible influence of grey extinction on distant SNe has been a topic of interest for the last twenty years, and a variety of different constraints have been established. In all cases those constraints have been formulated for models that are very different to ours and need to be reconsidered, as follows.

First, non-zero reddening is present to some degree in all models of dust extinction and has been used as a constraint by many authors \citep[e.g.][]{1999ApJ...525..583A,2002A&A...384....1G}. Unfortunately we do not know the extinction characteristics for the dust within the dense gas clouds under consideration here, nor do we know the details of their internal structure, so we cannot derive firm quantitative constraints. However, as shown in \S4.1 (particularly figure \ref{figure:reddening}) the ratio of total-to-selective extinction for dense clouds is much greater than that of the constituent material, so that reddening becomes a much less useful constraint. A combination of large dust grains located within dense, highly opaque clouds would yield almost completely grey extinction and would be very difficult to exclude on the basis of colour. 

Secondly, the absorption of starlight by intergalactic dust is expected to lead to a diffuse far-IR background as the dust reradiates the energy at longer wavelengths \citep[][]{2000ApJ...532...28A}. However, for the cold, dense gas that would make up the sort of objects we consider, the dust particles are likely to be predominantly solid H$_2$ snowflakes \citep[][]{walkerwardle2019}, with optical absorption cross-sections that are negligible compared to their scattering cross-sections \citep[][]{kettwich2015}. Consequently the level of far-IR background associated with a given level of optical extinction is decreased by orders of magnitude and this constraint becomes ineffective.

Thirdly, dust particles scatter X-rays through substantial angles, and the lack of extended (arcminute) haloes around distant quasars has been used to set limits on the amount of intergalactic dust \citep[][]{2006ApJ...651...41P}. However, X-rays scattered by dust that is located within dense gas clouds will be subject to photoelectric absorption by that gas, causing strong suppression of the scattering halo. Photoelectric absorption declines rapidly with increasing photon energy, so the suppression would be much weaker at high energies, but unfortunately most of the photon flux emitted by X-ray sources is usually to be found at low energies. Consequently the extinction constraints that can be derived from X-ray scattering haloes are also much weaker when that extinction is associated with dense gas clouds.

Finally, we are aware of one type of constraint on extinction that does not lose its effectiveness when the dust is in dense gas clouds: comparison between luminosity distances and angular diameter distances \citep[e.g.][]{2009ApJ...696.1727M}. In a transparent universe, luminosity distance and angular diameter distance differ only by a factor $(1+z)^q$ -- where the index $q$ depends on the type of luminosity under consideration (e.g. $q=2$ for bolometric luminosity) -- thus providing a direct test for transparency. The resulting constraints depend strongly on how the question is framed: (cosmological) model-dependent analyses indicate small extinction coefficients \citep[e.g. $\la\, 0.02\;{\rm mag\;Gpc^{-1}}$, ][]{2018MNRAS.477L..75G}, whereas (cosmological) model-independent analysis is much less constraining \citep[$\la\, 0.2\;{\rm mag \; Gpc^{-1}}$, ][]{2015PhRvD..92l3539L}. Clearly it is the latter, looser constraint that is the relevant one when it is the cosmological model itself that is at issue, so we are unable to exclude extinction by dense clouds as the cause of the dimming of distant SNe Ia.

\section{Quadrupole test}
\label{section:quadrupole}
The extent to which objects along the line of sight contribute to quasar variability has been a controversial subject. To help clarify the issue a number of tests have already been undertaken, with results that in the main favour a microlensing interpretation: chromaticity of the variability statistics \citep{2003MNRAS.344..492H}; lack of a cosmological time dilation \citep{2010MNRAS.405.1940H}; statistical symmetry in the light curves \citep[][]{1996MNRAS.278..787H}; and, the relationship between variations in the continuum and broad emission lines \citep{2011MNRAS.415.2744H}. Despite these successes the microlensing interpretation of quasar variability is not widely accepted --- perhaps because the framing of these tests relies, to some degree, on our imperfect understanding of quasar structure and evolution. 

Here we propose a new test --- one that is kinematic in nature and is therefore less model dependent than the tests which have been implemented to date. The timescale of the extrinsic variability -- be that gravitational or gas lensing or extinction -- depends on the effective transverse velocity of the line of sight, which is a weighted sum of the transverse velocities of the source, lens and observer. Now the transverse projection of the observer's three-dimensional velocity varies systematically over the sky, being zero in the direction parallel (or anti-parallel) to the observer motion, and is maximised at 90 degrees to that axis. Hence, there should be two poles of slowest variation on the sky and a band of fastest variation halfway between the poles; that is the test we are proposing, which we call `quadrupole' after its dominant pattern (as higher multipoles might be present depending on which quantity -- {\it e.g.}, timescale, variation rate {\it etc.} -- is actually measured). The influence of the observer's velocity on microlensing event rates has long been recognised \citep[e.g.][]{2011ApJ...738...96M,2016ApJ...832...46M,2020MNRAS.495..544N}, but to our knowledge it has not previously been understood to offer a test for an extrinsic origin of quasar variability.

The measured dipole component of the cosmic microwave background anisotropy corresponds to a speed of  $V_o=370\,\mathrm{km}\,\mathrm{s}^{-1}$ \citep{2020A&A...641A...1P} with respect to the CMB. Interpreted as a peculiar velocity this speed is unremarkable \citep{2013AJ....146...86T} and is widely interpreted as predominantly kinematic in origin \citep{2014A&A...571A..27P}, although uncomfortable discrepancies have been found for the amplitude of a related signal in the number counts of radio \citep{2019PhRvD.100f3501S} and infrared sources \citep{2021ApJ...908L..51S}. The peculiar velocities of individual lenses and sources are not known, but they are not expected to align with the observer velocity so the quadrupole pattern due to the observer's motion will be superimposed on a monopole. Moreover the quadrupole is small compared to the monopole, as we now show.

The expression for the effective velocity is \citep{1986A&A...166...36K}:
\begin{eqnarray}\label{veff}
\mathbf{v}_\mathrm{eff}=\mathbf{v}_l-\left(\frac{1+z_l}{1+z_s}\right)\frac{D_{l}}{D_{s}}\mathbf{v}_s-\frac{D_{ls}}{D_{s}}\mathbf{v}_o,
\end{eqnarray}
where $\mathbf{v}_o$ refers to the transverse (to the line of sight) component of the observer three-dimensional velocity such that
\begin{eqnarray}
v_o=V_o\sin\theta,
\end{eqnarray}
with $\theta$ being the angle the line of sight makes with the observer's motion. Assuming the first two terms in~(\ref{veff}) are drawn from a gaussian distribution with zero mean, so is their combination and the scaled effective velocity squared has a non-central $\chi^2$ distribution with two degrees of freedom. The mean of that distribution is
\begin{eqnarray}
\left\langle v_\mathrm{eff}^2\right\rangle_{v_{s,l}}=2\sigma^2+\left(\frac{D_{ls}}{D_{s}}\right)^2v_o^2,
\end{eqnarray}
where $\sigma^2$ is the (per spatial component) variance of the first two terms in equation~(\ref{veff}):
\begin{eqnarray}
\sigma^2=\sigma_l^2+\left(\frac{1+z_l}{1+z_s}\right)^2\left(\frac{D_{l}}{D_{s}}\right)^2\sigma_s^2.
\end{eqnarray}
The source velocity dispersion ought to be lower as the sources are seen at earlier stages in the development of cosmic structure, when the peculiar velocities are expected to be smaller, and because quasars are located near the centres of massive galaxies. The effect of the source velocity variance, $\sigma_s^2$, is itself reduced by redshift and lever-arm ratios and will be neglected in the following.

The relative amplitude of the quadrupole effect
\begin{eqnarray}
\eta\equiv\frac{\Delta\langle v_\mathrm{eff}\rangle_\mathrm{v_{s,l}}}{\overline{\langle v_\mathrm{eff}\rangle_\mathrm{v_{s,l}}}}\simeq\left(\frac{D_{ls}}{D_{s}}\right)^2\frac{V_o^2}{4\sigma^2}
\end{eqnarray}
depends explicitly on the position of the lenses along the line of sight and the appropriate effective value would depend on the effect responsible for variability. We assume that gravitational lensing, rather than extinction, is the variability mechanism of interest --- as is expected for high column-density clouds. Lensing is most effective when the critical density $\Sigma_{cr}(z_l, z_s)$ is lowest, which, at moderate source redshifts, occurs roughly halfway between the source and the observer. Thus the factor in parentheses above is close to $1/2$, yielding 
\begin{eqnarray}
\left\langle\eta\right\rangle_{l.o.s.}\simeq\frac{V_o^2}{16\sigma^2}
\end{eqnarray}
as our estimate. A more accurate estimate would require detailed numerical modelling.

For a (one-dimensional) lens velocity dispersion of $\sigma\sim400\,\mathrm{km}\,\mathrm{s}^{-1}$, as appropriate in the field, the relative amplitude of the variation across the sky is therefore small: $\eta\sim0.05$, requiring a precision of better than one percent in different sky directions. With the variations due to lensing being stochastic, the measured timescale itself is a random variable and the required precision can only be achieved by averaging over many sources. The variation in the timescale is related to those in the lens position along the line of sight, intrinsic spatial autocorrelation scale of the illumination pattern and relative velocity. Because lensing is ineffective for lenses close to the source and observer, this filter selects lenses at a particular distance (about halfway to the source) with moderate variation thereof. The intrinsic autocorrelation scale depends on the parameters of the lens, so the contribution of this effect into the variation will be modest if there is a preferred range of cloud parameters in the Universe. In this case the variation will be dominated by that of the effective velocity given by
\begin{eqnarray}
\frac{\left\langle(\Delta v_\mathrm{eff}^2\right\rangle}{\langle v_\mathrm{eff}^2\rangle}=\frac{\sqrt{1+(D_{ls}/D_{s})^2v_o^2/\sigma^2}}{1+(D_{ls}/D_{s})^2 v_o^2/2\sigma^2}\simeq 1
\end{eqnarray}
implying averaging variability timescales in $10^4$s sources for the required accuracy or $\sim$ per cent. This is likely a lower limit because it assumes that variability arising from any mechanism other than gravitational lensing can be neglected. On the other hand, detection could be aided by demanding that the anisotropy is aligned with that of the cosmological dipole. Ultimately, however, the required level of averaging will be determined empirically as a number of sources necessary to make the average stable from one patch of the sky to an adjacent one at the required level.

The above estimate suggests that the currently available quasar light curve data, such as the Southern Stripe 82 sample of the SDSS DR7 \citep{2012ApJ...753..106M}, are inadequate for the proposed test. However, high quality photometry of over a million quasars \citep{2020arXiv201202036G}, distributed all over the sky, will be made available in the next data release of the \emph{Gaia} mission; these data have the potential for a highly significant detection of the timescale-quadrupole, or else a rejection of the nanolensing interpretation of quasar variability. We caution that the peculiar -- and, most importantly, position-dependent -- time sampling of \emph{Gaia} targets will require careful accounting of the influence of that sampling on the derived timescale estimates.

\section{Summary and Conclusions}
\label{section:conclusions}
We have described the effects of a cosmological population of compact gas clouds on the radiation from distant sources. Gas clouds affect light in three different ways: through extinction, gas refraction and gravitational lensing, leading to a wide variety of possible light curves. The general character of the light curves can be anticipated for any circumstances and we demonstrated how to classify them according to the specified cloud properties. In the particular case of a gaussian column-density profile we showed the effect of a cloud of fixed mass but with central column-densities ranging over five orders of magnitude, including examples of weak lensing, strong gas lensing, and strong gravitational lensing --- for both transparent and highly opaque clouds.

Despite the potential importance of gas lensing -- which can substantially increase the lensing cross-section in some circumstances -- we find that neutral gas refraction has little influence on observations of quasars and type Ia supernovae at maximum light (or later). This result stands in contrast to \citet{draine1998}, and \citet{2001ApJ...547..207R}, who studied lensing of stars by gas clouds in and around the Galaxy. The origin of that difference is the much bigger sources that we consider --- which, at $R_s\sim (1-3)\times 10^{15}\;{\rm cm}$, are large compared to the gas lensing curvature radius ($\ell_\lambda\simeq 3.6\times 10^{13}\;{\rm cm}$), which in turn is large compared to the radii of dwarf stars ($R_s\sim 10^{11}\;{\rm cm}$). The key point is that we require a cloud size $r_0<\ell_\lambda$ for gas lensing to be stronger than gravitational lensing, so if $R_s \gg \ell_\lambda$ then that gas lens cannot greatly magnify the source.

For clouds with high central column-densities ($\Sigma_0\ga 1\;{\rm g\,cm^{-2}}$ for low-opacity clouds; $\Sigma_0\ga 10\;{\rm g\,cm^{-2}}$ for high-opacity clouds) our simulations yield quasar light-curves that are largely independent of that column --- all being similar to the point-mass case. This similarity is unsurprising because when gravitational lensing is important the dominant images form outside the cloud itself, so the details of its structure (and the associated gas refraction and extinction) have little influence. In this circumstance the mass of the individual lenses and their contribution to the mean cosmological density ($\Omega_l$) are the key variables that determine the appearance of the light-curves.

We find that high mass clouds ($ 10^{-4}\;{\rm M_\odot}\ll M \la 10^{-2}\;{\rm M_\odot}$) yield a great deal of large-amplitude quasar variability on timescales of a few years, and a high cosmological density of such objects can be excluded --- as originally pointed out by \cite{1993A&A...279....1S}. At the low end of our studied cloud mass range, below $10^{-5}\;{\rm M_\odot}$, we find only small variations in the magnification due to gravitational lensing, for sources as large as quasars; a high density of such clouds is permitted by the data, but cannot explain the variations that are seen. And between these extremes there is a narrow range of cloud mass, $M\sim 10^{-4.5\pm0.5}\;{\rm M_\odot}$, for which gravitational lensing by a population of gas clouds can explain much of the observed quasar variability, given a suitable mean cosmological density in such objects. 
All of these mass estimates would change systematically if we were to repeat our simulations using a quasar optical size that differs from our adopted value of $R_s=3\times 10^{15}\,{\rm cm}$, with the scaling being $M\propto R_s^2$ (for $R_s \sim 3\times 10^{15}\,{\rm cm}$).

If the material transparency is high the ``sweet spot'' -- wherein lensing can explain the observed variability -- applies to clouds across the full range of central column-densities that we studied ($0.1\;{\rm g\,cm^{-2}}\;\la\; \Sigma_0\;\la\; 1000\;{\rm g\,cm^{-2}}$). But if the material transparency is as low as in the local ISM then gaussian clouds that have low central columns ($\Sigma_0\la 1\;{\rm g\,cm^{-2}}$) are excluded because they introduce unacceptably large extinction fluctuations. Indeed such low-column, low-transparency gaussian clouds are in conflict with the quasar data for the full range of cloud masses studied in this paper.

The quasar variability data exclude large-amplitude extinction fluctuations but say little about the mean extinction. It is standard practice to use reddening as a gauge of extinction; however, we showed that the extinction introduced by dense gas clouds is very nearly colour-neutral --- even when the material opacity is strongly chromatic. This situation arises because of saturation: the large column-densities under consideration are totally opaque across a wide range of wavelengths, and the overall level of extinction of a source just reflects the fraction of its solid angle that is obscured by these opaque regions. As has been pointed out previously \citep[][]{1999ApJ...512L..19A}, the possible presence of grey extinction introduces ambiguity into the interpretation of data on standard candles such as type Ia supernovae \citep[e.g.][]{1998AJ....116.1009R,1998Natur.391...51P}. In this regard the dense gas clouds we have studied are particularly insidious because their total-to-selective extinction ratio is naturally very high indeed, and consequently colour is ineffective at discriminating between geometry and extinction. We have also noted how two other proposed constraints on intergalactic dust -- based on X-ray scattering \citep[e.g.][]{2006ApJ...651...41P}, and on the properties of the cosmic infrared background \citep[][]{2000ApJ...532...28A} -- are similarly ineffective. A direct test of cosmic transparency is available by comparing luminosity distances with angular diameter distances; however, when formulated in a framework that is independent of the cosmological model the resulting constraint is at present quite weak \citep[][]{2015PhRvD..92l3539L}.

Finally, we described a new statistical test which can differentiate between intrinsic and extrinsic origins of quasar variability. Extrinsic variations depend on the relative transverse velocity between the source, lens and observer, and the observer contribution to that velocity varies systematically over the sky. Because it relies only on kinematics, our test does not require the lenses to be gas clouds. The expected form of the variation over the sky is a quadrupolar pattern in the distribution of the average rate of external variations. With the source and lens velocity being randomly distributed, the amplitude of this variation is second order in the observer velocity and it is further suppressed by the lever-arm ratio that enters into the effective velocity. For gravitational lensing this results in an estimated amplitude of $\eta\sim0.05$, implying that the average variation timescales in different patches of the sky need to be measured to better than one percent. This is only possible by averaging timescales from a large number of light curves -- at least $10^4$ are needed to overcome the dispersion in the effective velocity itself. The currently available compilations of light curves are inadequate for this task, but the situation will improve when per-epoch \emph{Gaia} photometry of one million quasars becomes available. The complex sampling pattern of \emph{Gaia}, varying systematically across the sky, must be precisely accounted for when undertaking the proposed test.

\section*{Data availability}
The data underlying this article will be shared on reasonable request to the corresponding author.

\section*{Acknowledgements}
Helpful feedback on the quadrupole test was provided by Rudy Schild, Geraint Lewis and Vikram Ravi, and on the extinction of SNe Ia by Dominik Schwarz --- all much appreciated. Thanks also to Geraint for comments on the manuscript, and to the referee, Mike Hawkins, for some good suggestions for improvements. This research made use of Astropy,\footnote{http://www.astropy.org} a community-developed core Python package for Astronomy \citep{astropy:2013, astropy:2018}.

\bibliographystyle{mnras}
\bibliography{nanolensing} 

\begin{thebibliography}{}
\makeatletter
\relax
\def\mn@urlcharsother{\let\do\@makeother \do\$\do\&\do\#\do\^\do\_\do\%\do\~}
\def\mn@doi{\begingroup\mn@urlcharsother \@ifnextchar [ {\mn@doi@}
  {\mn@doi@[]}}
\def\mn@doi@[#1]#2{\def\@tempa{#1}\ifx\@tempa\@empty \href
  {http://dx.doi.org/#2} {doi:#2}\else \href {http://dx.doi.org/#2} {#1}\fi
  \endgroup}
\def\mn@eprint#1#2{\mn@eprint@#1:#2::\@nil}
\def\mn@eprint@arXiv#1{\href {http://arxiv.org/abs/#1} {{\tt arXiv:#1}}}
\def\mn@eprint@dblp#1{\href {http://dblp.uni-trier.de/rec/bibtex/#1.xml}
  {dblp:#1}}
\def\mn@eprint@#1:#2:#3:#4\@nil{\def\@tempa {#1}\def\@tempb {#2}\def\@tempc
  {#3}\ifx \@tempc \@empty \let \@tempc \@tempb \let \@tempb \@tempa \fi \ifx
  \@tempb \@empty \def\@tempb {arXiv}\fi \@ifundefined
  {mn@eprint@\@tempb}{\@tempb:\@tempc}{\expandafter \expandafter \csname
  mn@eprint@\@tempb\endcsname \expandafter{\@tempc}}}

\bibitem[\protect\citeauthoryear{{Aguirre}}{{Aguirre}}{1999a}]{1999ApJ...512L..19A}
{Aguirre} A.~N.,  1999a, \mn@doi [\apjl] {10.1086/311862}, \href
  {https://ui.adsabs.harvard.edu/abs/1999ApJ...512L..19A} {512, L19}

\bibitem[\protect\citeauthoryear{{Aguirre}}{{Aguirre}}{1999b}]{1999ApJ...525..583A}
{Aguirre} A.,  1999b, \mn@doi [\apj] {10.1086/307945}, \href
  {https://ui.adsabs.harvard.edu/abs/1999ApJ...525..583A} {525, 583}

\bibitem[\protect\citeauthoryear{{Aguirre} \& {Haiman}}{{Aguirre} \&
  {Haiman}}{2000}]{2000ApJ...532...28A}
{Aguirre} A.,  {Haiman} Z.,  2000, \mn@doi [\apj] {10.1086/308557}, \href
  {https://ui.adsabs.harvard.edu/abs/2000ApJ...532...28A} {532, 28}

\bibitem[\protect\citeauthoryear{{Alcock} et~al.,}{{Alcock}
  et~al.}{2000}]{2000ApJ...542..281A}
{Alcock} C.,  et~al., 2000, \mn@doi [\apj] {10.1086/309512}, \href
  {https://ui.adsabs.harvard.edu/abs/2000ApJ...542..281A} {542, 281}

\bibitem[\protect\citeauthoryear{{Astropy Collaboration} et~al.,}{{Astropy
  Collaboration} et~al.}{2013}]{astropy:2013}
{Astropy Collaboration} et~al., 2013, \mn@doi [\aap]
  {10.1051/0004-6361/201322068}, \href
  {http://adsabs.harvard.edu/abs/2013A%26A...558A..33A} {558, A33}

\bibitem[\protect\citeauthoryear{{Astropy Collaboration} et~al.,}{{Astropy
  Collaboration} et~al.}{2018}]{astropy:2018}
{Astropy Collaboration} et~al., 2018, \mn@doi [\aj] {10.3847/1538-3881/aabc4f},
  \href {https://ui.adsabs.harvard.edu/abs/2018AJ....156..123A} {156, 123}

\bibitem[\protect\citeauthoryear{{Blackburne}, {Pooley}, {Rappaport}  \&
  {Schechter}}{{Blackburne} et~al.}{2011}]{2011ApJ...729...34B}
{Blackburne} J.~A.,  {Pooley} D.,  {Rappaport} S.,   {Schechter} P.~L.,  2011,
  \mn@doi [\apj] {10.1088/0004-637X/729/1/34}, \href
  {https://ui.adsabs.harvard.edu/abs/2011ApJ...729...34B} {729, 34}

\bibitem[\protect\citeauthoryear{{Blandford} \& {McKee}}{{Blandford} \&
  {McKee}}{1982}]{1982ApJ...255..419B}
{Blandford} R.~D.,  {McKee} C.~F.,  1982, \mn@doi [\apj] {10.1086/159843},
  \href {https://ui.adsabs.harvard.edu/abs/1982ApJ...255..419B} {255, 419}

\bibitem[\protect\citeauthoryear{{Bohlin}, {Savage}  \& {Drake}}{{Bohlin}
  et~al.}{1978}]{bohlin1978}
{Bohlin} R.~C.,  {Savage} B.~D.,   {Drake} J.~F.,  1978, \mn@doi [\apj]
  {10.1086/156357}, \href
  {https://ui.adsabs.harvard.edu/abs/1978ApJ...224..132B} {224, 132}

\bibitem[\protect\citeauthoryear{{Bolatto}, {Wolfire}  \& {Leroy}}{{Bolatto}
  et~al.}{2013}]{2013ARA&A..51..207B}
{Bolatto} A.~D.,  {Wolfire} M.,   {Leroy} A.~K.,  2013, \mn@doi [\araa]
  {10.1146/annurev-astro-082812-140944}, \href
  {https://ui.adsabs.harvard.edu/abs/2013ARA&A..51..207B} {51, 207}

\bibitem[\protect\citeauthoryear{{Boveia} \& {Doglioni}}{{Boveia} \&
  {Doglioni}}{2018}]{2018ARNPS..68..429B}
{Boveia} A.,  {Doglioni} C.,  2018, \mn@doi [Annual Review of Nuclear and
  Particle Science] {10.1146/annurev-nucl-101917-021008}, \href
  {https://ui.adsabs.harvard.edu/abs/2018ARNPS..68..429B} {68, 429}

\bibitem[\protect\citeauthoryear{{Cardelli}, {Clayton}  \& {Mathis}}{{Cardelli}
  et~al.}{1989}]{1989ApJ...345..245C}
{Cardelli} J.~A.,  {Clayton} G.~C.,   {Mathis} J.~S.,  1989, \mn@doi [\apj]
  {10.1086/167900}, \href
  {https://ui.adsabs.harvard.edu/abs/1989ApJ...345..245C} {345, 245}

\bibitem[\protect\citeauthoryear{{Carr} \& {K{\"u}hnel}}{{Carr} \&
  {K{\"u}hnel}}{2020}]{2020ARNPS..70..355C}
{Carr} B.,  {K{\"u}hnel} F.,  2020, \mn@doi [Annual Review of Nuclear and
  Particle Science] {10.1146/annurev-nucl-050520-125911}, \href
  {https://ui.adsabs.harvard.edu/abs/2020ARNPS..70..355C} {70, 355}

\bibitem[\protect\citeauthoryear{{Carr}, {Kohri}, {Sendouda}  \&
  {Yokoyama}}{{Carr} et~al.}{2021}]{2021RPPh...84k6902C}
{Carr} B.,  {Kohri} K.,  {Sendouda} Y.,   {Yokoyama} J.,  2021, \mn@doi
  [Reports on Progress in Physics] {10.1088/1361-6633/ac1e31}, \href
  {https://ui.adsabs.harvard.edu/abs/2021RPPh...84k6902C} {84, 116902}

\bibitem[\protect\citeauthoryear{{Cen} \& {Ostriker}}{{Cen} \&
  {Ostriker}}{1999}]{1999ApJ...514....1C}
{Cen} R.,  {Ostriker} J.~P.,  1999, \mn@doi [\apj] {10.1086/306949}, \href
  {https://ui.adsabs.harvard.edu/abs/1999ApJ...514....1C} {514, 1}

\bibitem[\protect\citeauthoryear{{Clegg}, {Fey}  \& {Lazio}}{{Clegg}
  et~al.}{1998}]{1998ApJ...496..253C}
{Clegg} A.~W.,  {Fey} A.~L.,   {Lazio} T. J.~W.,  1998, \mn@doi [\apj]
  {10.1086/305344}, \href
  {https://ui.adsabs.harvard.edu/abs/1998ApJ...496..253C} {496, 253}

\bibitem[\protect\citeauthoryear{{Combes} \& {Pfenniger}}{{Combes} \&
  {Pfenniger}}{1997}]{1997A&A...327..453C}
{Combes} F.,  {Pfenniger} D.,  1997, \aap, \href
  {https://ui.adsabs.harvard.edu/abs/1997A&A...327..453C} {327, 453}

\bibitem[\protect\citeauthoryear{{Cornachione}, {Morgan}, {Millon}, {Bentz},
  {Courbin}, {Bonvin}  \& {Falco}}{{Cornachione}
  et~al.}{2020}]{2020ApJ...895..125C}
{Cornachione} M.~A.,  {Morgan} C.~W.,  {Millon} M.,  {Bentz} M.~C.,  {Courbin}
  F.,  {Bonvin} V.,   {Falco} E.~E.,  2020, \mn@doi [\apj]
  {10.3847/1538-4357/ab557a}, \href
  {https://ui.adsabs.harvard.edu/abs/2020ApJ...895..125C} {895, 125}

\bibitem[\protect\citeauthoryear{{Dai}, {Kochanek}, {Chartas}, {Koz{\l}owski},
  {Morgan}, {Garmire}  \& {Agol}}{{Dai} et~al.}{2010}]{2010ApJ...709..278D}
{Dai} X.,  {Kochanek} C.~S.,  {Chartas} G.,  {Koz{\l}owski} S.,  {Morgan}
  C.~W.,  {Garmire} G.,   {Agol} E.,  2010, \mn@doi [\apj]
  {10.1088/0004-637X/709/1/278}, \href
  {https://ui.adsabs.harvard.edu/abs/2010ApJ...709..278D} {709, 278}

\bibitem[\protect\citeauthoryear{{Dalgarno} \& {Williams}}{{Dalgarno} \&
  {Williams}}{1962}]{1962ApJ...136..690D}
{Dalgarno} A.,  {Williams} D.~A.,  1962, \mn@doi [\apj] {10.1086/147428}, \href
  {https://ui.adsabs.harvard.edu/abs/1962ApJ...136..690D} {136, 690}

\bibitem[\protect\citeauthoryear{{Draine}}{{Draine}}{1998}]{draine1998}
{Draine} B.~T.,  1998, \mn@doi [\apjl] {10.1086/311751}, \href
  {https://ui.adsabs.harvard.edu/abs/1998ApJ...509L..41D} {509, L41}

\bibitem[\protect\citeauthoryear{{Drake} \& {Cook}}{{Drake} \&
  {Cook}}{2003}]{2003ApJ...589..281D}
{Drake} A.~J.,  {Cook} K.~H.,  2003, \mn@doi [\apj] {10.1086/374640}, \href
  {https://ui.adsabs.harvard.edu/abs/2003ApJ...589..281D} {589, 281}

\bibitem[\protect\citeauthoryear{{Edelson} et~al.,}{{Edelson}
  et~al.}{2015}]{2015ApJ...806..129E}
{Edelson} R.,  et~al., 2015, \mn@doi [\apj] {10.1088/0004-637X/806/1/129},
  \href {https://ui.adsabs.harvard.edu/abs/2015ApJ...806..129E} {806, 129}

\bibitem[\protect\citeauthoryear{{Edelson} et~al.,}{{Edelson}
  et~al.}{2017}]{2017ApJ...840...41E}
{Edelson} R.,  et~al., 2017, \mn@doi [\apj] {10.3847/1538-4357/aa6890}, \href
  {https://ui.adsabs.harvard.edu/abs/2017ApJ...840...41E} {840, 41}

\bibitem[\protect\citeauthoryear{{Fausnaugh} et~al.,}{{Fausnaugh}
  et~al.}{2016}]{2016ApJ...821...56F}
{Fausnaugh} M.~M.,  et~al., 2016, \mn@doi [\apj] {10.3847/0004-637X/821/1/56},
  \href {https://ui.adsabs.harvard.edu/abs/2016ApJ...821...56F} {821, 56}

\bibitem[\protect\citeauthoryear{{Feng}}{{Feng}}{2010}]{2010ARA&A..48..495F}
{Feng} J.~L.,  2010, \mn@doi [\araa] {10.1146/annurev-astro-082708-101659},
  \href {https://ui.adsabs.harvard.edu/abs/2010ARA&A..48..495F} {48, 495}

\bibitem[\protect\citeauthoryear{{Fukugita}, {Hogan}  \& {Peebles}}{{Fukugita}
  et~al.}{1998}]{1998ApJ...503..518F}
{Fukugita} M.,  {Hogan} C.~J.,   {Peebles} P.~J.~E.,  1998, \mn@doi [\apj]
  {10.1086/306025}, \href
  {https://ui.adsabs.harvard.edu/abs/1998ApJ...503..518F} {503, 518}

\bibitem[\protect\citeauthoryear{{Gaia Collaboration} et~al.,}{{Gaia
  Collaboration} et~al.}{2020}]{2020arXiv201202036G}
{Gaia Collaboration} et~al., 2020, arXiv e-prints, \href
  {https://ui.adsabs.harvard.edu/abs/2020arXiv201202036G} {p. arXiv:2012.02036}

\bibitem[\protect\citeauthoryear{{Gerhard} \& {Silk}}{{Gerhard} \&
  {Silk}}{1996}]{1996ApJ...472...34G}
{Gerhard} O.,  {Silk} J.,  1996, \mn@doi [\apj] {10.1086/178039}, \href
  {https://ui.adsabs.harvard.edu/abs/1996ApJ...472...34G} {472, 34}

\bibitem[\protect\citeauthoryear{{Goobar}, {Bergstr{\"o}m}  \&
  {M{\"o}rtsell}}{{Goobar} et~al.}{2002}]{2002A&A...384....1G}
{Goobar} A.,  {Bergstr{\"o}m} L.,   {M{\"o}rtsell} E.,  2002, \mn@doi [\aap]
  {10.1051/0004-6361:20020002}, \href
  {https://ui.adsabs.harvard.edu/abs/2002A&A...384....1G} {384, 1}

\bibitem[\protect\citeauthoryear{{Goobar}, {Dhawan}  \& {Scolnic}}{{Goobar}
  et~al.}{2018}]{2018MNRAS.477L..75G}
{Goobar} A.,  {Dhawan} S.,   {Scolnic} D.,  2018, \mn@doi [\mnras]
  {10.1093/mnrasl/sly053}, \href
  {https://ui.adsabs.harvard.edu/abs/2018MNRAS.477L..75G} {477, L75}

\bibitem[\protect\citeauthoryear{{Hawkins}}{{Hawkins}}{1993}]{1993Natur.366..242H}
{Hawkins} M.~R.~S.,  1993, \mn@doi [\nat] {10.1038/366242a0}, \href
  {https://ui.adsabs.harvard.edu/abs/1993Natur.366..242H} {366, 242}

\bibitem[\protect\citeauthoryear{{Hawkins}}{{Hawkins}}{1996}]{1996MNRAS.278..787H}
{Hawkins} M.~R.~S.,  1996, \mn@doi [\mnras] {10.1093/mnras/278.3.787}, \href
  {https://ui.adsabs.harvard.edu/abs/1996MNRAS.278..787H} {278, 787}

\bibitem[\protect\citeauthoryear{{Hawkins}}{{Hawkins}}{2003}]{2003MNRAS.344..492H}
{Hawkins} M.~R.~S.,  2003, \mn@doi [\mnras] {10.1046/j.1365-8711.2003.06828.x},
  \href {https://ui.adsabs.harvard.edu/abs/2003MNRAS.344..492H} {344, 492}

\bibitem[\protect\citeauthoryear{{Hawkins}}{{Hawkins}}{2010}]{2010MNRAS.405.1940H}
{Hawkins} M.~R.~S.,  2010, \mn@doi [\mnras] {10.1111/j.1365-2966.2010.16581.x},
  \href {https://ui.adsabs.harvard.edu/abs/2010MNRAS.405.1940H} {405, 1940}

\bibitem[\protect\citeauthoryear{{Hawkins}}{{Hawkins}}{2011}]{2011MNRAS.415.2744H}
{Hawkins} M.~R.~S.,  2011, \mn@doi [\mnras] {10.1111/j.1365-2966.2011.18890.x},
  \href {https://ui.adsabs.harvard.edu/abs/2011MNRAS.415.2744H} {415, 2744}

\bibitem[\protect\citeauthoryear{{Hawkins}}{{Hawkins}}{2020a}]{2020A&A...633A.107H}
{Hawkins} M.~R.~S.,  2020a, \mn@doi [\aap] {10.1051/0004-6361/201936462}, \href
  {https://ui.adsabs.harvard.edu/abs/2020A&A...633A.107H} {633, A107}

\bibitem[\protect\citeauthoryear{{Hawkins}}{{Hawkins}}{2020b}]{2020A&A...643A..10H}
{Hawkins} M.~R.~S.,  2020b, \mn@doi [\aap] {10.1051/0004-6361/202038670}, \href
  {https://ui.adsabs.harvard.edu/abs/2020A&A...643A..10H} {643, A10}

\bibitem[\protect\citeauthoryear{{Hawkins} \& {Veron}}{{Hawkins} \&
  {Veron}}{1993}]{1993MNRAS.260..202H}
{Hawkins} M.~R.~S.,  {Veron} P.,  1993, \mn@doi [\mnras]
  {10.1093/mnras/260.1.202}, \href
  {https://ui.adsabs.harvard.edu/abs/1993MNRAS.260..202H} {260, 202}

\bibitem[\protect\citeauthoryear{{Helbig}}{{Helbig}}{2020}]{2020OJAp....3E...1H}
{Helbig} P.,  2020, \mn@doi [The Open Journal of Astrophysics]
  {10.21105/astro.1912.12269}, \href
  {https://ui.adsabs.harvard.edu/abs/2020OJAp....3E...1H} {3, 1}

\bibitem[\protect\citeauthoryear{{Henriksen} \& {Widrow}}{{Henriksen} \&
  {Widrow}}{1995}]{1995ApJ...441...70H}
{Henriksen} R.~N.,  {Widrow} L.~M.,  1995, \mn@doi [\apj] {10.1086/175336},
  \href {https://ui.adsabs.harvard.edu/abs/1995ApJ...441...70H} {441, 70}

\bibitem[\protect\citeauthoryear{{Homayouni} et~al.,}{{Homayouni}
  et~al.}{2019}]{2019ApJ...880..126H}
{Homayouni} Y.,  et~al., 2019, \mn@doi [\apj] {10.3847/1538-4357/ab2638}, \href
  {https://ui.adsabs.harvard.edu/abs/2019ApJ...880..126H} {880, 126}

\bibitem[\protect\citeauthoryear{{Huchra}, {Gorenstein}, {Kent}, {Shapiro},
  {Smith}, {Horine}  \& {Perley}}{{Huchra} et~al.}{1985}]{1985AJ.....90..691H}
{Huchra} J.,  {Gorenstein} M.,  {Kent} S.,  {Shapiro} I.,  {Smith} G.,
  {Horine} E.,   {Perley} R.,  1985, \mn@doi [\aj] {10.1086/113777}, \href
  {https://ui.adsabs.harvard.edu/abs/1985AJ.....90..691H} {90, 691}

\bibitem[\protect\citeauthoryear{{Jiang} et~al.,}{{Jiang}
  et~al.}{2017}]{2017ApJ...836..186J}
{Jiang} Y.-F.,  et~al., 2017, \mn@doi [\apj] {10.3847/1538-4357/aa5b91}, \href
  {https://ui.adsabs.harvard.edu/abs/2017ApJ...836..186J} {836, 186}

\bibitem[\protect\citeauthoryear{{Jim{\'e}nez-Vicente}, {Mediavilla},
  {Kochanek}  \& {Mu{\~n}oz}}{{Jim{\'e}nez-Vicente}
  et~al.}{2015}]{2015ApJ...799..149J}
{Jim{\'e}nez-Vicente} J.,  {Mediavilla} E.,  {Kochanek} C.~S.,   {Mu{\~n}oz}
  J.~A.,  2015, \mn@doi [\apj] {10.1088/0004-637X/799/2/149}, \href
  {https://ui.adsabs.harvard.edu/abs/2015ApJ...799..149J} {799, 149}

\bibitem[\protect\citeauthoryear{{Kayser}, {Refsdal}  \& {Stabell}}{{Kayser}
  et~al.}{1986}]{1986A&A...166...36K}
{Kayser} R.,  {Refsdal} S.,   {Stabell} R.,  1986, \aap, \href
  {https://ui.adsabs.harvard.edu/abs/1986A&A...166...36K} {166, 36}

\bibitem[\protect\citeauthoryear{{Kerins}, {Binney}  \& {Silk}}{{Kerins}
  et~al.}{2002}]{2002MNRAS.332L..29K}
{Kerins} E.,  {Binney} J.,   {Silk} J.,  2002, \mn@doi [\mnras]
  {10.1046/j.1365-8711.2002.05463.x}, \href
  {https://ui.adsabs.harvard.edu/abs/2002MNRAS.332L..29K} {332, L29}

\bibitem[\protect\citeauthoryear{{Kettwich}, {Anderson}, {Walker}  \&
  {Tuntsov}}{{Kettwich} et~al.}{2015}]{kettwich2015}
{Kettwich} S.~C.,  {Anderson} D.~T.,  {Walker} M.~A.,   {Tuntsov} A.~V.,  2015,
  \mn@doi [\mnras] {10.1093/mnras/stv691}, \href
  {https://ui.adsabs.harvard.edu/abs/2015MNRAS.450.1032K} {450, 1032}

\bibitem[\protect\citeauthoryear{{Kochanek}}{{Kochanek}}{2004}]{2004ApJ...605...58K}
{Kochanek} C.~S.,  2004, \mn@doi [\apj] {10.1086/382180}, \href
  {https://ui.adsabs.harvard.edu/abs/2004ApJ...605...58K} {605, 58}

\bibitem[\protect\citeauthoryear{{Liao}, {Avgoustidis}  \& {Li}}{{Liao}
  et~al.}{2015}]{2015PhRvD..92l3539L}
{Liao} K.,  {Avgoustidis} A.,   {Li} Z.,  2015, \mn@doi [\prd]
  {10.1103/PhysRevD.92.123539}, \href
  {https://ui.adsabs.harvard.edu/abs/2015PhRvD..92l3539L} {92, 123539}

\bibitem[\protect\citeauthoryear{{MacLeod} et~al.,}{{MacLeod}
  et~al.}{2012}]{2012ApJ...753..106M}
{MacLeod} C.~L.,  et~al., 2012, \mn@doi [\apj] {10.1088/0004-637X/753/2/106},
  \href {https://ui.adsabs.harvard.edu/abs/2012ApJ...753..106M} {753, 106}

\bibitem[\protect\citeauthoryear{{Macquart} et~al.,}{{Macquart}
  et~al.}{2020}]{2020Natur.581..391M}
{Macquart} J.~P.,  et~al., 2020, \mn@doi [\nat] {10.1038/s41586-020-2300-2},
  \href {https://ui.adsabs.harvard.edu/abs/2020Natur.581..391M} {581, 391}

\bibitem[\protect\citeauthoryear{{Mediavilla}, {Jim{\'e}nez-Vicente},
  {Mu{\~n}oz}  \& {Battaner}}{{Mediavilla} et~al.}{2016}]{2016ApJ...832...46M}
{Mediavilla} E.,  {Jim{\'e}nez-Vicente} J.,  {Mu{\~n}oz} J.~A.,   {Battaner}
  E.,  2016, \mn@doi [\apj] {10.3847/0004-637X/832/1/46}, \href
  {https://ui.adsabs.harvard.edu/abs/2016ApJ...832...46M} {832, 46}

\bibitem[\protect\citeauthoryear{{Metcalf} \& {Silk}}{{Metcalf} \&
  {Silk}}{1999}]{1999ApJ...519L...1M}
{Metcalf} R.~B.,  {Silk} J.,  1999, \mn@doi [\apjl] {10.1086/312086}, \href
  {https://ui.adsabs.harvard.edu/abs/1999ApJ...519L...1M} {519, L1}

\bibitem[\protect\citeauthoryear{{More}, {Bovy}  \& {Hogg}}{{More}
  et~al.}{2009}]{2009ApJ...696.1727M}
{More} S.,  {Bovy} J.,   {Hogg} D.~W.,  2009, \mn@doi [\apj]
  {10.1088/0004-637X/696/2/1727}, \href
  {https://ui.adsabs.harvard.edu/abs/2009ApJ...696.1727M} {696, 1727}

\bibitem[\protect\citeauthoryear{{Morgan}, {Kochanek}, {Morgan}  \&
  {Falco}}{{Morgan} et~al.}{2010}]{2010ApJ...712.1129M}
{Morgan} C.~W.,  {Kochanek} C.~S.,  {Morgan} N.~D.,   {Falco} E.~E.,  2010,
  \mn@doi [\apj] {10.1088/0004-637X/712/2/1129}, \href
  {https://ui.adsabs.harvard.edu/abs/2010ApJ...712.1129M} {712, 1129}

\bibitem[\protect\citeauthoryear{{Morgan}, {Hyer}, {Bonvin}, {Mosquera},
  {Cornachione}, {Courbin}, {Kochanek}  \& {Falco}}{{Morgan}
  et~al.}{2018}]{2018ApJ...869..106M}
{Morgan} C.~W.,  {Hyer} G.~E.,  {Bonvin} V.,  {Mosquera} A.~M.,  {Cornachione}
  M.,  {Courbin} F.,  {Kochanek} C.~S.,   {Falco} E.~E.,  2018, \mn@doi [\apj]
  {10.3847/1538-4357/aaed3e}, \href
  {https://ui.adsabs.harvard.edu/abs/2018ApJ...869..106M} {869, 106}

\bibitem[\protect\citeauthoryear{{Mortonson}, {Schechter}  \&
  {Wambsganss}}{{Mortonson} et~al.}{2005}]{2005ApJ...628..594M}
{Mortonson} M.~J.,  {Schechter} P.~L.,   {Wambsganss} J.,  2005, \mn@doi [\apj]
  {10.1086/431195}, \href
  {https://ui.adsabs.harvard.edu/abs/2005ApJ...628..594M} {628, 594}

\bibitem[\protect\citeauthoryear{{Mosquera} \& {Kochanek}}{{Mosquera} \&
  {Kochanek}}{2011}]{2011ApJ...738...96M}
{Mosquera} A.~M.,  {Kochanek} C.~S.,  2011, \mn@doi [\apj]
  {10.1088/0004-637X/738/1/96}, \href
  {https://ui.adsabs.harvard.edu/abs/2011ApJ...738...96M} {738, 96}

\bibitem[\protect\citeauthoryear{{Mudd} et~al.,}{{Mudd}
  et~al.}{2018}]{2018ApJ...862..123M}
{Mudd} D.,  et~al., 2018, \mn@doi [\apj] {10.3847/1538-4357/aac9bb}, \href
  {https://ui.adsabs.harvard.edu/abs/2018ApJ...862..123M} {862, 123}

\bibitem[\protect\citeauthoryear{{Neira}, {Anguita}  \& {Vernardos}}{{Neira}
  et~al.}{2020}]{2020MNRAS.495..544N}
{Neira} F.,  {Anguita} T.,   {Vernardos} G.,  2020, \mn@doi [\mnras]
  {10.1093/mnras/staa1208}, \href
  {https://ui.adsabs.harvard.edu/abs/2020MNRAS.495..544N} {495, 544}

\bibitem[\protect\citeauthoryear{{Paczynski}}{{Paczynski}}{1986}]{1986ApJ...304....1P}
{Paczynski} B.,  1986, \mn@doi [\apj] {10.1086/164140}, \href
  {https://ui.adsabs.harvard.edu/abs/1986ApJ...304....1P} {304, 1}

\bibitem[\protect\citeauthoryear{{Perlmutter} et~al.,}{{Perlmutter}
  et~al.}{1998}]{1998Natur.391...51P}
{Perlmutter} S.,  et~al., 1998, \mn@doi [\nat] {10.1038/34124}, \href
  {https://ui.adsabs.harvard.edu/abs/1998Natur.391...51P} {391, 51}

\bibitem[\protect\citeauthoryear{{Petric}, {Telis}, {Paerels}  \&
  {Helfand}}{{Petric} et~al.}{2006}]{2006ApJ...651...41P}
{Petric} A.,  {Telis} G.~A.,  {Paerels} F.,   {Helfand} D.~J.,  2006, \mn@doi
  [\apj] {10.1086/507667}, \href
  {https://ui.adsabs.harvard.edu/abs/2006ApJ...651...41P} {651, 41}

\bibitem[\protect\citeauthoryear{{Pfenniger} \& {Combes}}{{Pfenniger} \&
  {Combes}}{1994}]{1994A&A...285...94P}
{Pfenniger} D.,  {Combes} F.,  1994, \aap, \href
  {https://ui.adsabs.harvard.edu/abs/1994A&A...285...94P} {285, 94}

\bibitem[\protect\citeauthoryear{{Pfenniger}, {Combes}  \&
  {Martinet}}{{Pfenniger} et~al.}{1994}]{1994A&A...285...79P}
{Pfenniger} D.,  {Combes} F.,   {Martinet} L.,  1994, \aap, \href
  {https://ui.adsabs.harvard.edu/abs/1994A&A...285...79P} {285, 79}

\bibitem[\protect\citeauthoryear{{Planck Collaboration} et~al.,}{{Planck
  Collaboration} et~al.}{2014}]{2014A&A...571A..27P}
{Planck Collaboration} et~al., 2014, \mn@doi [\aap]
  {10.1051/0004-6361/201321556}, \href
  {https://ui.adsabs.harvard.edu/abs/2014A&A...571A..27P} {571, A27}

\bibitem[\protect\citeauthoryear{{Planck Collaboration} et~al.,}{{Planck
  Collaboration} et~al.}{2016}]{2016A&A...594A..13P}
{Planck Collaboration} et~al., 2016, \mn@doi [\aap]
  {10.1051/0004-6361/201525830}, \href
  {https://ui.adsabs.harvard.edu/abs/2016A&A...594A..13P} {594, A13}

\bibitem[\protect\citeauthoryear{{Planck Collaboration} et~al.,}{{Planck
  Collaboration} et~al.}{2020}]{2020A&A...641A...1P}
{Planck Collaboration} et~al., 2020, \mn@doi [\aap]
  {10.1051/0004-6361/201833880}, \href
  {https://ui.adsabs.harvard.edu/abs/2020A&A...641A...1P} {641, A1}

\bibitem[\protect\citeauthoryear{{Poindexter} \& {Kochanek}}{{Poindexter} \&
  {Kochanek}}{2010a}]{2010ApJ...712..658P}
{Poindexter} S.,  {Kochanek} C.~S.,  2010a, \mn@doi [\apj]
  {10.1088/0004-637X/712/1/658}, \href
  {https://ui.adsabs.harvard.edu/abs/2010ApJ...712..658P} {712, 658}

\bibitem[\protect\citeauthoryear{{Poindexter} \& {Kochanek}}{{Poindexter} \&
  {Kochanek}}{2010b}]{2010ApJ...712..668P}
{Poindexter} S.,  {Kochanek} C.~S.,  2010b, \mn@doi [\apj]
  {10.1088/0004-637X/712/1/668}, \href
  {https://ui.adsabs.harvard.edu/abs/2010ApJ...712..668P} {712, 668}

\bibitem[\protect\citeauthoryear{{Pooley}, {Blackburne}, {Rappaport}  \&
  {Schechter}}{{Pooley} et~al.}{2007}]{2007ApJ...661...19P}
{Pooley} D.,  {Blackburne} J.~A.,  {Rappaport} S.,   {Schechter} P.~L.,  2007,
  \mn@doi [\apj] {10.1086/512115}, \href
  {https://ui.adsabs.harvard.edu/abs/2007ApJ...661...19P} {661, 19}

\bibitem[\protect\citeauthoryear{{Pringle}}{{Pringle}}{1981}]{1981ARA&A..19..137P}
{Pringle} J.~E.,  1981, \mn@doi [\araa] {10.1146/annurev.aa.19.090181.001033},
  \href {https://ui.adsabs.harvard.edu/abs/1981ARA&A..19..137P} {19, 137}

\bibitem[\protect\citeauthoryear{{Rafikov} \& {Draine}}{{Rafikov} \&
  {Draine}}{2001}]{2001ApJ...547..207R}
{Rafikov} R.~R.,  {Draine} B.~T.,  2001, \mn@doi [\apj] {10.1086/318355}, \href
  {https://ui.adsabs.harvard.edu/abs/2001ApJ...547..207R} {547, 207}

\bibitem[\protect\citeauthoryear{{Riess} et~al.,}{{Riess}
  et~al.}{1998}]{1998AJ....116.1009R}
{Riess} A.~G.,  et~al., 1998, \mn@doi [\aj] {10.1086/300499}, \href
  {https://ui.adsabs.harvard.edu/abs/1998AJ....116.1009R} {116, 1009}

\bibitem[\protect\citeauthoryear{{Riess} et~al.,}{{Riess}
  et~al.}{2004}]{2004ApJ...607..665R}
{Riess} A.~G.,  et~al., 2004, \mn@doi [\apj] {10.1086/383612}, \href
  {https://ui.adsabs.harvard.edu/abs/2004ApJ...607..665R} {607, 665}

\bibitem[\protect\citeauthoryear{{Schechter} \& {Wambsganss}}{{Schechter} \&
  {Wambsganss}}{2004}]{2004IAUS..220..103S}
{Schechter} P.~L.,  {Wambsganss} J.,  2004, in {Ryder} S.,  {Pisano} D.,
  {Walker} M.,   {Freeman} K.,  eds,  Vol. 220, Dark Matter in Galaxies. p.~103
  (\mn@eprint {arXiv} {astro-ph/0309163})

\bibitem[\protect\citeauthoryear{{Schild}}{{Schild}}{1996}]{1996ApJ...464..125S}
{Schild} R.~E.,  1996, \mn@doi [\apj] {10.1086/177304}, \href
  {https://ui.adsabs.harvard.edu/abs/1996ApJ...464..125S} {464, 125}

\bibitem[\protect\citeauthoryear{{Schneider}}{{Schneider}}{1993}]{1993A&A...279....1S}
{Schneider} P.,  1993, \aap, \href
  {https://ui.adsabs.harvard.edu/abs/1993A&A...279....1S} {279, 1}

\bibitem[\protect\citeauthoryear{{Schneider}, {Ehlers}  \& {Falco}}{{Schneider}
  et~al.}{1992}]{sef1992}
{Schneider} P.,  {Ehlers} J.,   {Falco} E.~E.,  1992, {Gravitational Lenses},
  \mn@doi{10.1007/978-3-662-03758-4.
}

\bibitem[\protect\citeauthoryear{{Secrest}, {von Hausegger}, {Rameez},
  {Mohayaee}, {Sarkar}  \& {Colin}}{{Secrest}
  et~al.}{2021}]{2021ApJ...908L..51S}
{Secrest} N.~J.,  {von Hausegger} S.,  {Rameez} M.,  {Mohayaee} R.,  {Sarkar}
  S.,   {Colin} J.,  2021, \mn@doi [\apjl] {10.3847/2041-8213/abdd40}, \href
  {https://ui.adsabs.harvard.edu/abs/2021ApJ...908L..51S} {908, L51}

\bibitem[\protect\citeauthoryear{{Shakura} \& {Sunyaev}}{{Shakura} \&
  {Sunyaev}}{1973}]{1973A&A....24..337S}
{Shakura} N.~I.,  {Sunyaev} R.~A.,  1973, \aap, \href
  {https://ui.adsabs.harvard.edu/abs/1973A&A....24..337S} {500, 33}

\bibitem[\protect\citeauthoryear{{Singal}}{{Singal}}{2019}]{2019PhRvD.100f3501S}
{Singal} A.~K.,  2019, \mn@doi [\prd] {10.1103/PhysRevD.100.063501}, \href
  {https://ui.adsabs.harvard.edu/abs/2019PhRvD.100f3501S} {100, 063501}

\bibitem[\protect\citeauthoryear{{Trimble}}{{Trimble}}{1987}]{1987ARA&A..25..425T}
{Trimble} V.,  1987, \mn@doi [\araa] {10.1146/annurev.aa.25.090187.002233},
  \href {https://ui.adsabs.harvard.edu/abs/1987ARA&A..25..425T} {25, 425}

\bibitem[\protect\citeauthoryear{{Tully} et~al.,}{{Tully}
  et~al.}{2013}]{2013AJ....146...86T}
{Tully} R.~B.,  et~al., 2013, \mn@doi [\aj] {10.1088/0004-6256/146/4/86}, \href
  {https://ui.adsabs.harvard.edu/abs/2013AJ....146...86T} {146, 86}

\bibitem[\protect\citeauthoryear{{Udalski}, {Zebrun}, {Szymanski}, {Kubiak},
  {Pietrzynski}, {Soszynski}  \& {Wozniak}}{{Udalski}
  et~al.}{2000}]{2000AcA....50....1U}
{Udalski} A.,  {Zebrun} K.,  {Szymanski} M.,  {Kubiak} M.,  {Pietrzynski} G.,
  {Soszynski} I.,   {Wozniak} P.,  2000, \actaa, \href
  {https://ui.adsabs.harvard.edu/abs/2000AcA....50....1U} {50, 1}

\bibitem[\protect\citeauthoryear{{Walker}}{{Walker}}{1999a}]{1999MNRAS.306..504W}
{Walker} M.~A.,  1999a, \mn@doi [\mnras] {10.1046/j.1365-8711.1999.02537.x},
  \href {https://ui.adsabs.harvard.edu/abs/1999MNRAS.306..504W} {306, 504}

\bibitem[\protect\citeauthoryear{{Walker}}{{Walker}}{1999b}]{1999MNRAS.308..551W}
{Walker} M.~A.,  1999b, \mn@doi [\mnras] {10.1046/j.1365-8711.1999.02814.x},
  \href {https://ui.adsabs.harvard.edu/abs/1999MNRAS.308..551W} {308, 551}

\bibitem[\protect\citeauthoryear{{Walker} \& {Wardle}}{{Walker} \&
  {Wardle}}{2019}]{walkerwardle2019}
{Walker} M.~A.,  {Wardle} M.~J.,  2019, \mn@doi [\apj]
  {10.3847/1538-4357/ab2987}, \href
  {https://ui.adsabs.harvard.edu/abs/2019ApJ...881...69W} {881, 69}

\bibitem[\protect\citeauthoryear{{Wechsler} \& {Tinker}}{{Wechsler} \&
  {Tinker}}{2018}]{2018ARA&A..56..435W}
{Wechsler} R.~H.,  {Tinker} J.~L.,  2018, \mn@doi [\araa]
  {10.1146/annurev-astro-081817-051756}, \href
  {https://ui.adsabs.harvard.edu/abs/2018ARA&A..56..435W} {56, 435}

\bibitem[\protect\citeauthoryear{{Wyithe}, {Webster}  \& {Turner}}{{Wyithe}
  et~al.}{2000}]{2000MNRAS.315...51W}
{Wyithe} J.~S.~B.,  {Webster} R.~L.,   {Turner} E.~L.,  2000, \mn@doi [\mnras]
  {10.1046/j.1365-8711.2000.03360.x}, \href
  {https://ui.adsabs.harvard.edu/abs/2000MNRAS.315...51W} {315, 51}

\bibitem[\protect\citeauthoryear{{Yu} et~al.,}{{Yu}
  et~al.}{2020}]{2020ApJS..246...16Y}
{Yu} Z.,  et~al., 2020, \mn@doi [\apjs] {10.3847/1538-4365/ab5e7a}, \href
  {https://ui.adsabs.harvard.edu/abs/2020ApJS..246...16Y} {246, 16}

\bibitem[\protect\citeauthoryear{{Zackrisson} \& {Bergvall}}{{Zackrisson} \&
  {Bergvall}}{2003}]{2003A&A...399...23Z}
{Zackrisson} E.,  {Bergvall} N.,  2003, \mn@doi [\aap]
  {10.1051/0004-6361:20021762}, \href
  {https://ui.adsabs.harvard.edu/abs/2003A&A...399...23Z} {399, 23}

\bibitem[\protect\citeauthoryear{{Zackrisson}, {Bergvall}, {Marquart}  \&
  {Helbig}}{{Zackrisson} et~al.}{2003}]{2003A&A...408...17Z}
{Zackrisson} E.,  {Bergvall} N.,  {Marquart} T.,   {Helbig} P.,  2003, \mn@doi
  [\aap] {10.1051/0004-6361:20030895}, \href
  {https://ui.adsabs.harvard.edu/abs/2003A&A...408...17Z} {408, 17}

\bibitem[\protect\citeauthoryear{{Zumalac{\'a}rregui} \&
  {Seljak}}{{Zumalac{\'a}rregui} \& {Seljak}}{2018}]{2018PhRvL.121n1101Z}
{Zumalac{\'a}rregui} M.,  {Seljak} U.,  2018, \mn@doi [\prl]
  {10.1103/PhysRevLett.121.141101}, \href
  {https://ui.adsabs.harvard.edu/abs/2018PhRvL.121n1101Z} {121, 141101}

\bibitem[\protect\citeauthoryear{{de Paolis}, {Ingrosso}, {Jetzer}  \&
  {Roncadelli}}{{de Paolis} et~al.}{1995}]{1995PhRvL..74...14D}
{de Paolis} F.,  {Ingrosso} G.,  {Jetzer} P.,   {Roncadelli} M.,  1995, \mn@doi
  [\prl] {10.1103/PhysRevLett.74.14}, \href
  {https://ui.adsabs.harvard.edu/abs/1995PhRvL..74...14D} {74, 14}

\bibitem[\protect\citeauthoryear{{de Swart}, {Bertone}  \& {van Dongen}}{{de
  Swart} et~al.}{2017}]{2017NatAs...1E..59D}
{de Swart} J.~G.,  {Bertone} G.,   {van Dongen} J.,  2017, \mn@doi [Nature
  Astronomy] {10.1038/s41550-017-0059}, \href
  {https://ui.adsabs.harvard.edu/abs/2017NatAs...1E..59D} {1, 0059}

\makeatother
\end{thebibliography}

\appendix
\section{Quasar optical continuum size}
Source size plays a critical r\^ole in the modelling of nanolensing variations because it determines the spatial smoothing scale of the flux (magnification) map. In the present paper we use a representative source radius of $R_s=3\times 10^{15}\,\mathrm{cm}$, which is three times larger than the value used by \cite{1993A&A...279....1S}. Our chosen value is close to that expected from standard, thin disc theory for a $10^9\,{\rm M}_\odot$ black hole accreting at a fair fraction of the Eddington rate \citep[e.g.][]{2010ApJ...712.1129M}. In this Appendix we describe three different ways of estimating the size of the optical continuum emission region in quasars. 

\subsection{Photometric estimate for standard accretion disc}
Models of thin accretion discs are simplest in the case where the radiation is optically-thick thermal emission, which is locally well approximated by the Planck function, and at large radii, $R$, the temperature, $T$, varies as a power-law: $T\propto R^{-3/4}$ \citep{1973A&A....24..337S,1981ARA&A..19..137P}. This functional form is expected to be a good approximation outside the hottest portion of the disc, which in turn is close to its inner edge at 1--6 gravitational radii (depending on the angular momentum of the hole), i.e. $10^{14.2-15}\,\mathrm{cm}\,\times (M_\mathrm{BH}/10^9\,{\rm M}_\odot$).

In the case of locally black-body emission and a power-law variation of temperature with radius, an observing frequency $\nu$ picks out a characteristic radius, $R_\nu$, such that
\begin{eqnarray}
\frac{h\nu}{k T}=\left(\frac{R}{R_\nu}\right)^{2/m}.
\end{eqnarray}
In terms of that radius, the luminosity spectral density, $L_\nu$, inferred by an observer at infinity, and polar angle $\theta$ relative to the disc normal, is given by 
\begin{eqnarray}\label{BBLuminosity}
L_\nu = 16\pi^2R_\nu^2 \,\cos\theta\,\frac{h\nu^3}{c^2}\;{\cal  I}(m),
\end{eqnarray}
where
\begin{eqnarray}
{\cal I}(m)\equiv \int_0^\infty\, \frac{x\,{\rm d}x}{\exp(x^{2/m})-1}\quad = \quad \frac{1}{2}\Gamma(1+m)\zeta(m),
\end{eqnarray}
with $\Gamma$ and $\zeta$ being the Gamma (factorial) Function and the Riemann Zeta Function, respectively. The integral converges for $1<m<\infty$. For a standard thin disk the appropriate power-law is $m=8/3$ and the integral evaluates to ${\cal I}(8/3)\simeq 2.576$.

It is important to note that equation (\ref{BBLuminosity}) does not explicitly depend on the mass of the black hole, its accretion rate, or the viscosity in the disc. Consequently the luminosity can be inferred from the measured source flux and redshift alone, within the adopted cosmological model, and thus we can immediately determine the corresponding value of $R_\nu$ --- up to a factor of order unity that is associated with the unknown disc inclination.

In \S3, for reasons of computational efficiency, we used a uniform surface-brightness source model. In detail the nanolensing flux variations do depend on the surface-brightness profile of the source, so there can be only an approximate equivalence between that description and an accretion disc model. However, it is known that the source size -- as gauged by the half-light radius -- is the most important factor in determining the magnification statistics \citep{2005ApJ...628..594M}. For a disc having a power-law temperature profile with $m=8/3$ the half-light radius is $R_{1/2}\simeq 2.436\,R_\nu$, whereas a uniform disc has a radius, $R_s$, that is $\sqrt{2}$ larger than its half-light radius, thus the radius of the equivalent uniform disc is $R_s\simeq 3.444\,R_\nu(m=8/3)$. 

Figure A1 shows the value of $R_s$ appropriate to the observed $B$-band magnitudes and redshifts of the \cite{1993MNRAS.260..202H} sample of quasars in the same (Planck) cosmological model that we used in \S3. In all cases we have taken a representative value of $\cos\theta=0.5$. From this figure we conclude that a suitable radius for a uniform disc model for this sample of sources is $R_s = 3\times 10^{15}\,{\rm cm}$.

In the case of other physical models that predict different power-law variations of temperature with radius, the equivalent radius of a uniform disc scales approximately as $R_s\propto m^2$ in the vicinity of $m=8/3$ (valid over the range $1.5 \la m \la 4.5$).

\begin{figure}
    \includegraphics[width=\columnwidth]{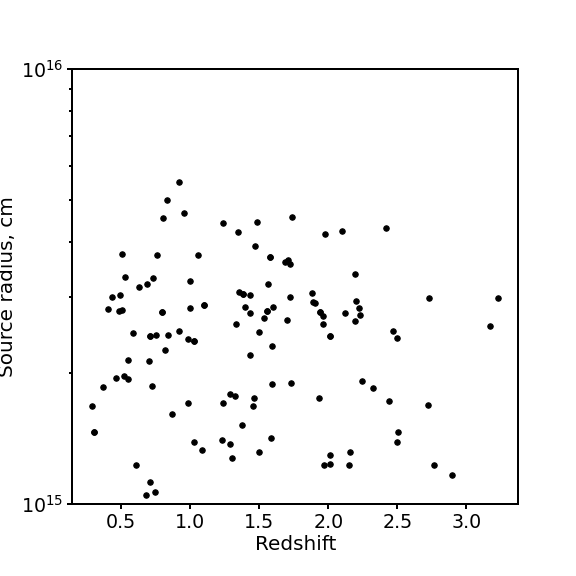}
    \caption{The radius of the equivalent disc of uniform surface-brightness having the same half-light radius as a locally black-body disc ($R_s = 3.444\,R_\nu(m=8/3)$), for each of the  quasars in the \citet{1993MNRAS.260..202H} sample. These values are computed using equation (\ref{BBLuminosity}), assuming no extinction and average inclination of the accretion disc to the line of sight ($\cos\theta=0.5$). All radii are for the observed $B$-band, which corresponds to progressively shorter wavelengths in the rest frame of the quasar at higher redshifts.}
    \label{figure:hvsizes}
\end{figure}

\subsection{Continuum reverberation mapping}
Reverberation mapping is a technique that has its origins in studies of the broad line region (BLR) of quasars. The physical picture is that the emission lines arise from photoionisation and should therefore vary in response to fluctuations in the UV continuum flux from an accretion disc. The temporal impulse response function of the BLR, as recorded by an observer far away, depends on its geometry, so photometric monitoring of both continuum and line fluxes can constrain the spatial distribution of the emission line clouds \citep[][]{1982ApJ...255..419B}.

Continuum reverberation mapping is a related idea: high energy photons are imagined to be absorbed by the accretion disc itself (e.g. because the disc is flared), and the absorbed power is subsequently reradiated in the form of lower energy photons. In this setting photometric monitoring in a variety of continuum passbands can constrain the geometry of the continuum emitting regions. A detailed physical description of this mechanism is currently lacking, but even so one can examine data to see whether they correspond well to the idea, and to establish quantitative constraints. 

Continuum reverberation mapping measurements of nearby active galactic nuclei in NGC~4151 and NGC~5548 (which have well-measured black hole masses) confirmed the predicted scaling of accretion disc size with wavelength (equation A1 with $m=8/3$), but returned lag amplitude about a factor of 3 higher than expected \citep{2016ApJ...821...56F, 2015ApJ...806..129E}. At the same time these studies called into question the validity of the temporal response model -- often referred to as the ``lamp post model'' -- underlying the method \citep{2017ApJ...840...41E}. A similar size discrepancy was reported from a statistical analysis of high signal-to-noise lag measurements for a large number of quasars in the Pan-STARRS data \citep{2017ApJ...836..186J}.

Recent continuum reverberation studies, however, have not supported the idea of large continuum sources. High quality light curves from the Sloan Digital Sky Survey \citep[nearly a hundred quasars,][]{2019ApJ...880..126H} and Dark Energy Survey \citep[over twenty quasars,][]{2018ApJ...862..123M, 2020ApJS..246...16Y} returned lag estimates -- hence source sizes -- that are fully consistent with the size of standard, thin accretion discs. Furthermore it was pointed out by \citet{2019ApJ...880..126H} that earlier statistical work, which focused on objects with high signal-to-noise lag measurements, was strongly biased towards large lags and therefore large disc sizes. In summary: at present there is no evidence from continuum reverberation mapping for quasar sizes larger than expected in the standard, thin accretion disc theory.

\subsection{Microlensing of multiply-imaged quasars}
Quasars that are gravitationally lensed by foreground galaxies can offer insights into the properties of the source. When there are multiple, distinct images one can obtain information about the instrinsic changes in source flux -- which, up to a lag, are the same for all images -- and also about the microlensing, which differs amongst the different paths. As noted earlier, the character of the microlensing signal is strongly affected by the size of the source.

\subsubsection{Flux ratio anomalies}
Smooth models of the gravitational potential of the lens galaxy can explain the positions of the lensed images, but the observed flux ratios often differ from the corresponding predictions and are thus termed ``anomalous'' \citep[e.g.][]{2004IAUS..220..103S}. In reality the lens galaxy potential has structure on small scales -- e.g. due to individual stars -- and so the observed flux ratio anomalies have a natural interpretation in terms of small-scale structure in the lensing magnification map.

In a study of ten quadruply imaged quasars \cite{2007ApJ...661...19P} found that the flux ratio anomalies are much smaller in the optical than in the X-ray. They interpreted this as the quasar optical emission region being larger than the X-ray emission region, so that the magnification map is more strongly smoothed in the optical. Using simulations \cite{2007ApJ...661...19P} determined  the optical source size to be one-tenth to one-third of the typical Einstein ring radius of the microlenses. These results were reinforced with an augmented sample of 12 quasars by \cite{2011ApJ...729...34B} --- who also find a source size scaling with wavelength that is much flatter, on average, than the standard accretion disc theory predicts.

In order to translate these results into an absolute source size one needs to know the typical Einstein ring radius of the microlenses, whereas the data themselves do not determine that radius. \cite{2007ApJ...661...19P} and \cite{2011ApJ...729...34B} assumed stellar mass microlenses, which led them to conclude that optical quasars are much larger than standard, thin accretion discs. If the dark matter is indeed an elementary particle then the assumption of stellar mass microlenses is reasonable. However, in this paper we are considering the possibility of a universe with a high density of sub-stellar objects, and in that context one also expects small-scale structure in the magnification map arising from nanolenses.

\subsubsection{Light-curve analysis}
The degeneracy just mentioned can in principle be lifted in studies of microlensing light-curves, because the temporal coordinate translates to a spatial scale through the effective transverse speed. However, that quantity is itself not known a priori and the best that can be done is to construct a probability distribution based on reasonable physical assumptions. 

One lensed quasar, Q2237$+$0305 \citep[][]{1985AJ.....90..691H}, has received a great deal of attention in the literature --- because of the uniquely low redshift of the lens galaxy, and the correspondingly short timescale of its microlensing variations. This lens galaxy is a massive spiral and all four images lie within one kpc of the centre, so we are seeing the quasar through a high surface density of stars. In this case one might imagine that we are on firm ground if we assume the microlenses to be of stellar mass. In fact \citet{2000MNRAS.315...51W} excluded the possibility that a significant fraction of the total surface density could be in planetary mass objects, on the grounds that the resulting microlensing rate would be much higher than observed. By the same token, however, it remains possible that the observed microlensing could include a significant contribution from planetary mass objects: even a small surface density in such objects could yield a higher event rate than the population of stellar microlenses. In connection with this possibility it is worth noting that a virialised dark halo of dense gas clouds is expected to have a core in its density profile \citep[][]{1999MNRAS.308..551W}, as a result of physical collisions, and should indeed make only a small contribution to the surface-density near the centre \emph{even if dense gas comprises all of the dark matter}.

Contemporary analyses of microlensing in Q2237$+$0305 utilise Bayesian inference to identify viable combinations of the source size, microlens mass and effective transverse speed. Early work \citep[e.g.][]{2004ApJ...605...58K} returned source sizes consistent with standard accretion discs, whereas recent results have favoured larger sources \citep[e.g.][]{2010ApJ...712..658P,2010ApJ...712..668P}. However, to date none of these analyses have considered the circumstance just mentioned, in which a small admixture of planetary mass objects contributes significantly to the observed microlensing signal. Consequently it is unclear what these studies have to say about the appropriate quasar size to use in the present paper.

In large part the analyses of microlensing in other multiply-imaged quasars have been fashioned after the pioneering Q2237$+$0305 studies \citep[notably the work of][]{2004ApJ...605...58K}, and there are many papers which favour quasar sizes significantly larger than standard accretion discs \citep[e.g.][]{2010ApJ...709..278D,2018ApJ...869..106M,2020ApJ...895..125C}. However, just like Q2237$+$0305, all of these studies are based on monomodal distributions of the microlens mass, whereas a bimodal distribution is required if the dark matter consists of planetary mass lumps.

\label{lastpage}
\end{document}